\documentclass[amsmath,amssymb, superscriptaddress, preprint]{revtex4-1}

\usepackage[T1]{fontenc}
\usepackage[ansinew]{inputenc}
\usepackage[english]{babel}
\usepackage{amsfonts}
\usepackage{array}
\usepackage{amssymb}
\usepackage{amsthm}
\usepackage{amsmath}
\usepackage{comment}
\usepackage{epsfig}
\usepackage{tabu}
\usepackage{subfig}
\usepackage{float}
\usepackage{epstopdf}
\usepackage{multirow}
\usepackage{graphicx}
\usepackage{eucal}
\usepackage{verbatim}
\usepackage[table]{xcolor}
\usepackage{caption}
\usepackage{bm}
\usepackage[pdftex,plainpages=false,colorlinks,hyperindex,bookmarksopen,linkcolor=black,citecolor=blue,urlcolor=blue]{hyperref}
\usepackage{braket}
\newcommand{\be}{\begin{eqnarray}}
\newcommand{\ee}{\end{eqnarray}}

\newcommand{\R}{\mathbb{R}}  %%%%% \R = \mathbb{R}.
\def\eg{{\it e.g. }} 
\def\ie{{\it i.e. }}

\theoremstyle{definition}
\newtheorem{definition}{Definition}[section]

\begin{document}

\title{Virasoro Algebra and Asymptotic Symmetries from Fractional Bosonic Strings}

	\author{Victor Alfonzo Diaz}
		\email{vadrienzo@gmail.com}
		\affiliation{Dipartimento di Fisica e Astronomia, Universit\`a di Bologna, Via Irnerio~46, 40126 Bologna, Italy.}
		\affiliation{ INFN, Sezione di Bologna, viale Berti Pichat 6/2, 40127 Bologna, Italy}

%	\date  {\today}%%{January 2016}

\begin{abstract}

The aim of this work is to further study the fractional bosonic string theory. In particular, we wrote the energy-momentum tensor in the fractional conformal gauge and study their symmetries. We introduced the Virasoro operators of all orders. In fact, we found the same $L_0 (\widetilde{L}_0)$ operator originally defined in the work of fractional bosonic string up to a shift transformation. 
Also, we compute the algebra of our Fractional Virasoro Operators, finding that the satifies the $Witt$ algebra. Lastly, we showed that in the boundary of our theory we recover the lost conservation law
associated to $\tau$-diffeomorphism, proving that we have Poincar\'e invariance at the boundary.
\end{abstract}

\maketitle
\section{Introduction}
Fractional calculus has played and still plays an important role in several fields, with tremendous applications in engineering and physics, chemistry, and biology. In particular, the formulation of the Lagrange formalism within Riemann-Liouville fractional derivatives, \eg see for example \cite{MainardiB, IC-AG-FM-ZAMP, IC-AG-FM-Bessel, AG-FCAA-2017, extra, AG-FM_MECC16, JMP-mio, Garra, Silvia-1, Vacaru} together with the study of variational problems with fractional derivatives \cite{ Ata, Almeida, Torres-book, Calcagni-rev, Calcagni-PRL} opened an interesting playground to re-formulate certain well known problems in physic, always in a context of classical field theory.

These new formalism led to the study of several works in the branch of gravity and cosmology~\cite{Cosmo-1,Cosmo-2}. In the discourse of re-formulation of well known theories, a natural and fundamental question arise: can String Theory be re-developed in the context of fractional variational problem? An affirmative answer was found in \cite{Diaz:2017hdd} with the introduction of Fractional Bosonic Strings. 

Fractional Bosonic Strings is formulated as a generalization of the Polyakov action in the framework of fractional calculus. The concept of a fractional time-like parameter is introduced which replaces the usual $\tau$-parameter in the World Sheet (WS). Using the Riemann-Liouville fractional integral we introduce a new measure in the usual WS action. This new extra factor led to the breaking of the diffeomorphism invariance in the time direction, reducing the number of symmetries in the theory. Although, highly non-linear equations of motion were found, it was showed that in the fractional conformal gauge they were solvable.

The purpose of this work is to further study the solution found in~\cite{Diaz:2017hdd}. In particular filling the gaps in the previous work. We delve on topics such as the conserved currents associated to the diffeomorphism invariance, computing explicitly the continuity equations. Together with the introduction the Fractional Virasoro Operators. 

The work is organized as follows: In Section \ref{Sec1} we reviewed the Fractional Bosonic Strings~\cite{Diaz:2017hdd}. We write the time-Fractional Polyakov action, together with the associated equations of motion. We introduce the energy-momentum and we compute continuity equations with their respective Noether currents. We define the \textit{generalized light-cone coordinates}, where for a flat auxiliary metric on the WS reduces to the usual light-cone coordinates. After properly discuss the energy-momentum tensor and the continuity equations, we move on to write the solution of the equation of motions. In particular, we  study the solution of the embedding map $\bm{X}(\tau,\sigma)$ for periodic boundary conditions in the fractional conformal gauge. Subsequently in Section \ref{Sec2}, we give a proper definitions of the Hamiltonian together with the operator performing rigid translations in the $\sigma$-direction $P_\sigma$. Leading us to an uni-vocal representation, in the fractional conformal gauge, of the Fractional Virasoro Operators of order $p$. Once with a proper definition of the Virasoro, we focus in Section \ref{Sec3} to the computation of the algebra satisfied by the $\alpha$'s oscillators and by transitivity in Section \ref{Sec4} the algebra satisfied by the Virasoro operators, which is in fact proven to satisfied the Witt algebra. An interesting result is presented in Section \ref{Sec5}, where it is shown that in the boundary of the WS we recover the time diffeomorphism invariance. We regain the lost homogeneous continuity equation associated to time diffeomorphism invariance, implying that in the asymptotic limit we recover the symmetry. It is not surprising to find that in the boundary of out theory we recover such symmetry. In fact, this is a well study process in gravity and gauge theories called asymptotic symmetries, \eg see~\cite{Bondi:1962px, Sachs:1962wk, Strominger:2013lka, Conde:2016csj, Brown:1986nw, Strominger:2017zoo}. Finally, we draw our conclusion. Appendix \ref{Appen:Prop} contains several properties of the function $\mathcal{E}_\nu^m(z)$ used during the development of the work. In the Appendix \ref{Appen:ContiEqu}, we checked the conituity equation related to $\sigma$-diffeomorphism invariance is satisfied. In the Appendix \ref{Appen:ViraSums} the Poisson brackets of the Fractional Virasoro Operators is explicit computed. Lastly as new interesting representation of the Virasoro generators, we present in Appendix \ref{Appen:AsympViraSums} the asymptotic form of the Fractional Virasoro Operators and we computed their algebra.

\section{Review of Fractional Bosonic String}\label{Sec1}
Let us begin by reviewing the Fractional Bosonic strings, briefly discussing the \textit{time-fractional Polyakov action} and the equation of motion. Then, proceed afterwards by computing the energy-momentum tensor and the continuity equations. Finally, showing the solution of the equation of motion in the fractional conformal gauge.

\subsection{Time-fractional Polyakov action}

The dynamic of a propagating string can be describe by the Polyakov action \cite{deser, brink, polyakov, lust, Zwiebach}, 
\begin{eqnarray}\label{Polyact}
S_{P}=-\frac{1}{4\pi\alpha'}\int_{\Sigma} d^2\bm{\sigma} \,\sqrt{-h}\,h^{ab}(\bm{\sigma})\,\partial_a \bm{X}(\bm{\sigma}) \cdot \partial_b \bm{X} (\bm{\sigma}) \, ,
\end{eqnarray}
where
\begin{eqnarray}
\bm{X} \, : \, \Sigma \, \longrightarrow \, \mathbb{R}^{1,d-1} \, , \quad \bm{\sigma}\,:\,(\tau , \, \sigma) \mapsto \bm{X} (\bm{\sigma}) = \{ X^\mu (\tau, \sigma)\, , \,\, \quad \mu = 0, 1, \ldots, d-1 \} \,.
\end{eqnarray}
A generalization of this action was introduced in the framework of fractional calculus~\cite{Diaz:2017hdd}, where the fractional action is given by
\begin{eqnarray}\label{PolyactFrac}
S^{\alpha, \beta} [h, \bm{X}] = -\frac{1}{4\pi\alpha'} \int _\Sigma d^2\bm{\sigma} \, v_{\alpha , \beta} (\bm{ \sigma}) 
\,\sqrt{-h}\,h^{ab}(\bm{\sigma})\,\partial_a \bm{X}(\bm{\sigma)} \cdot \partial_b \bm{X} (\bm{\sigma}) \, ,
\end{eqnarray}
with the extra function $v_{\alpha,\beta}(\bm{\sigma})$ coming from the Lebesgue-Stieltjes measure and $0 \leq \alpha , \beta \leq 1$.	In particular, we can choose the case in which
	\be \label{RL-time-measure}
	\Sigma = \Sigma _t = ( - \infty, t ) \, \times \,  [0, 2\pi] \, , \quad 
	v_{\alpha , \beta} (\bm{\sigma})  = \, _{-\infty} v_{\alpha , 1} (t; \tau, \sigma) = 
	\frac{(t - \tau)^{\alpha - 1}}{\Gamma (\alpha)} \, ,
	\ee 
with $0 < \alpha \leq 1$ and $t \in \R$. Therefore, in this specific example we can write the \textit{time-fractional Polyakov action} as
	\be \label{our-model}
	S_{\alpha}\equiv-\frac{1}{4\pi\alpha'\,\Gamma(\alpha)}\int_{0}^{2\pi}d\sigma\,\int_{-\infty}^{t}(t-\tau)^{\alpha-1}\,d\tau\left[\sqrt{-h}\,h^{ab}(\tau,\sigma)\,\partial_a \bm{X}(\tau,\sigma) \cdot \partial_b \bm{X} (\tau,\sigma) \right].
	\ee
In the limit  where $\alpha\rightarrow1$ and $t\rightarrow+\infty$ we can easily see that we recover the Polyakov action \eqref{Polyact}. It also is clear that these limits do not commute, \ie we must take the $\alpha$-limit before the $t$-limit, otherwise we encounter a divergence in the action. 

It is important to notice that as in the usual bosonic string theory we could ask if there exist other terms that one could add to \eqref{Polyact}. Our mainly discussion will be based on closed strings propagating in Mikowski space with no other background fields, therefore following the usual arguments given in~\cite{lust} the only terms compatible with Poincar\`e invariance in $d$-dimension and power counting renormalizability (with maximum two derivatives) of the two-dimensional theory are in the fractional generalized case
\be\label{Act1}
S_{1,\alpha}&\equiv&\frac{\lambda_1}{\Gamma(\alpha)}\int_{0}^{2\pi}d\sigma\,\int_{-\infty}^{t}(t-\tau)^{\alpha-1}\,d\tau\,\sqrt{-h}\,;\\
\label{Act2}
S_{2,\alpha}&\equiv&\frac{\lambda_2}{2\pi\Gamma(\alpha)}\int_{0}^{2\pi}d\sigma\,\int_{-\infty}^{t}(t-\tau)^{\alpha-1}\,d\tau\,\sqrt{-h}\,R^{(2)}\,,
\ee
where $\lambda_1,\lambda_2\in\mathbb{R}$, and $R^{(2)}$ is the Ricci scalar associated the induce metric in the WS. The first action \eqref{Act1} is the generalization of the cosmological constant term and the second action \eqref{Act2} is the generalization of the Gauss-Bonnet term. The addition of both terms to the Polyakov action breaks Weyl invariance and change the equation of motion for the induced metric $h_{ab}$, then being inconsistent with the result presented in~\cite{Diaz:2017hdd}. From now on, we will assume the vanishing of the cosmological constant ($\lambda_1=0$) and $\lambda_2=0$. The next step is to review the computation of the equations of motion.
 \subsection{Equations of motion} \label{sec-EOM}
 
As in every dynamical theory the equations of motion are found by taking the variation of the action and setting it to zero with suitable boundary conditions. Taking the variation of the action \eqref{PolyactFrac} with respect to each field thereby, we find
	\be \label{eq-motion-gen-1}
	\delta _{\bm{X}} S_\alpha = 0 \,\, \rightsquigarrow \,\, 
	\partial_a \left[ (t-\tau)^{\alpha-1} \, \sqrt{-h} \, h^{ab} \, \partial_b \bm{X} \right] = 0 \, ,
	\ee
	\be\label{eq-motion-gen-2}	
	\delta _{h} S_\alpha= 0 \,\, \rightsquigarrow \,\, 
	\partial_a \bm{X} \cdot\partial_b \bm{X} - \frac{1}{2}\,h_{ab}\,h^{cd}\,\partial_c \bm{X} \cdot \partial_d \bm{X} = 0 \, .
	\ee
	If we define the energy-momentum tensor of the WS theory in the usual way, \ie
	\be \label{stress-energy}
	T_{ab} \equiv - \frac{4 \pi}{\sqrt{-h}} \, \frac{\delta S_\alpha [h, \bm{X}]}{\delta h ^{a b}} = 
\frac{(t-\tau)^{\alpha-1}}{\alpha'\,\Gamma(\alpha)}\left[ \partial_a \bm{X} \cdot\partial_b \bm{X} -\frac{1}{2}\,h_{ab}\,h^{cd}\,\partial_c \bm{X} \cdot\partial_d \bm{X}  \right]\, ,
	\ee
	and defining the improved d'Alembert operator $\Box _{\alpha}$ as
	\be 
	\Box _{\alpha} \phi \equiv \frac{1}{(t-\tau)^{\alpha-1}\, \sqrt{-h}} \, 
	\partial_a \left[(t-\tau)^{\alpha-1}\, \sqrt{-h} \, h^{ab} \, \partial_b \phi  \right] \, ,
	\ee
	then Eq. \eqref{eq-motion-gen-1} and \eqref{eq-motion-gen-2} read,
	\be 
	T_{ab} = 0 \, ,
	\ee
	\be
	\Box _{\alpha} \bm{X} = 0 \, .
	\ee

Through the use of our symmetries we can reduce the degrees of freedom (d.o.f.) of the auxiliary metric $h_{ab}$. It is well known that in 2 dimension the metric $h_{ab}$ has 3 d.o.f, using the reparametrization in the $\sigma$-direction and by means of the Weyl invariance, we can remove two of them. This leaves us with only one d.o.f for the metric. Therefore, we can choose the \textit{fractional conformal gauge} and write 
\begin{eqnarray} \label{our-metric}
h_{ab}=\left(
\begin{array}{cc}
-1 & 0\\
 0 &[f^2(\tau,\sigma)]^{\alpha-1}
\end{array}
\right),
\end{eqnarray}
with $f(\tau,\sigma)$ an arbitrary function and in the limit $\alpha\rightarrow1$, we recover the usual flat metric.  
Making use of metric \eqref{our-metric} the equations of motion can be written as
\be \label{our-EOM-sym-1}
(\alpha-1)f^{2\alpha-2}\left[\frac{\dot{f}}{f}-\frac{1}{t-\tau}\right]\,\dot{\bm{X}}+(\alpha-1)\,\frac{f'}{f}\,\bm{X}'-\bm{X}''+f^{2\alpha-2}\,\ddot{\bm{X}} 
=0 \, ,
\ee
\be \label{our-EOM-sym-2}
|| f^{\alpha-1}\,\dot{\bm{X}}\pm \bm{X}' ||^2 = 0 \, ,
\ee
where the prime represents the derivative with respect to $\sigma$, the dot is understood as the derivative with respect to $\tau$ and $|| \bm{a} ||^2 := \bm{a} \cdot \bm{a}$.
\subsection{Energy-momentum tensor}\label{EneMomentTenso}

In the fractional conformal gauge, we already saw that the auxiliary metric can be written as in \eqref{our-metric}. Therefore, the non-zero Christoffel symbols are given by
\be
\Gamma^{0}_{11}=(\alpha-1)\,\frac{\dot{f}}{f}\,f^{2\alpha-2}\quad;\quad\Gamma^{1}_{01}=(\alpha-1)\,\frac{\dot{f}}{f}\quad;\quad\Gamma^{1}_{11}=(\alpha-1)\,\frac{f'}{f},
\ee
where for $f(\tau,\sigma)=1$, we recover the trivial result of a flat space $\Gamma^{a}_{bc}=0$. 

In general Noether theorem states that every symmetry of the action correspond to a conservation law. In particular for diffeomorphism invariance the conservation law is given by
\be
\nabla^a T_{ab}=0\quad\Rightarrow\quad \partial^aT_{ab}-h^{ac}\Gamma^d_{ca}T_{db}-h^{ac}\Gamma^d_{cb}T_{da}=0\,.
\ee
In the case of the \textit{Fractional Bosonic String} we have broken explicitly the diffeomorphism invariance in the $\tau$-direction. Therefore, we do not have all the conservation laws. In this theory we have
\be\label{conservationlaw}
\nabla^a T_{a1}=0\quad\mbox{and}\quad \nabla^a T_{a0}=(1-\alpha)\,\partial_0\Big[\ln(t-\tau)\Big]\,T_{00}\,,
\ee
where one can see that in the limit $\alpha\rightarrow1$ we restore the conservation law corresponding to the $\tau$-diffeomorphism invariance. For simplicity, we define
\be
g_{\alpha}(\tau,\sigma)\equiv(1-\alpha)\,\partial_0\Big[\ln(t-\tau)\Big]\,T_{00}\,.
\ee
Using in the fractional conformal gauge the explicit expression of $T_{ab}$ \eqref{stress-energy}, we can write
\be\label{continuiyEqFracBos}
\nabla^a T_{a1}=0\quad&\Rightarrow&\quad -\partial_0 T_{01}+\partial_1 T_{00}- (\alpha-1)\,\frac{\dot{f}}{f}\,T_{01}=0\,;\\\label{continuiyEqFracBos2}
\nabla^a T_{a0}=g_{\alpha}(\tau,\sigma)\quad&\Rightarrow&\quad -\partial_0 T_{11}+\partial_1 T_{01}- (\alpha-1)\,\frac{f'}{f}\,T_{01}=f^{2\alpha-2}\,g_{\alpha}(\tau,\sigma)\,.
\ee
Here, we have used
\be\label{T01}
T_{01}=T_{10}&=& \frac{(t-\tau)^{\alpha-1}}{2\,\alpha'\,\Gamma(\alpha)}\left(2\,\dot{\bm{X}}\cdot \bm{X}'\right)\,;\\\label{T00}
T_{00}=f^{2-2\alpha}T_{11}&=& \frac{(t-\tau)^{\alpha-1}}{2\,\alpha'\,\Gamma(\alpha)}\left(||\dot{\bm{X}}||^2+f^{2-2\alpha}\,||\bm{X}'||^{2}\right)\,.
\ee

For the later purpose it is useful to introduce the \textit{generalized light-cone coordinates}: 
$$d\sigma^+=d\tau+f^{\alpha-1}\,d\sigma\quad,\quad d\sigma^+=d\tau-f^{\alpha-1}\,d\sigma\,\,\Rightarrow ds^2=-d\sigma^+ d\sigma^-\,.$$ 
Therefore, we can define 
\be\label{T++T--}
T_{++}\equiv\frac{1}{2}\left(f^{\alpha-1}\,T_{00}+T_{01}\right)\quad\mbox{and} \quad T_{--}\equiv\frac{1}{2}\left(f^{\alpha-1}\,T_{00}-T_{01}\right).
\ee

In this new set of coordinates one can write
\be\label{partial+-}
\partial_0=\partial_+ +\partial_-\quad\mbox{and}\quad\partial_1=f^{\alpha-1}\left(\partial_+-\partial_-\right).
\ee
%Replacing \eqref{T++T--} and \eqref{partial+-} in \eqref{continuiyEqFracBos} and \eqref{continuiyEqFracBos2} we find
%\be
%\partial_+ T_{--}-\partial_- T_{++}-\frac{(1-\alpha)}{2f}\left[ \left( \dot{f}+f'\,f^{1-\alpha}\right)\,T_{++}- \left(\dot{f}- f'\,f^{1-\alpha}\right)\,T_{--} \right]&=&0\,;\\\
%\partial_+ T_{--}+\partial_- T_{++}-\frac{(1-\alpha)}{2f}\left[ \left( \dot{f}+f'\,f^{1-\alpha}\right)\,T_{++}+ \left(\dot{f}- f'\,f^{1-\alpha}\right)\,T_{--} \right]&=&\frac{f^{\alpha-1}g_{\alpha}(\tau,\sigma)}{2}\,.
%\ee
For the case $f(\tau,\sigma)=1$ we go back to the usual light-cone coordinates and one can write
\be\label{contiT++T--}
\partial_+ T_{--}-\partial_- T_{++}=0\quad\mbox{and}\quad\partial_+ T_{--}+\partial_- T_{++}=\frac{(1-\alpha)}{2}\partial_0\Big[\ln(t-\tau)\Big]\,\Big(T_{++} +T_{--}\Big).
\ee
These equations will be used as an important cross-check for the consistency of our solution in the case of $f(\tau,\sigma)=1$.
 
\subsection{Solution of the equation of motion}\label{Sol.Eq.Motion}

It has been shown in~\cite{Diaz:2017hdd}, that for $f(\tau,\sigma)=1$, the embedding map of the string $\bm{X}(\tau,\sigma)$ can be expanded by
\begin{equation} \label{eq-sol}
\begin{split}
 \bm{X} (z, \sigma) = \bm{X}_0 -\sqrt{2\alpha'}\, \frac{ z^{2 \nu}}{2 \nu} \, \bm{\alpha} _0
  - i \, \sqrt{\frac{\alpha'}{2}} \sum _{m \neq 0}  \Big( \frac{\bm{\alpha} _m}{m} \, e^{i \, m\, \sigma} +  \frac{\widetilde{\bm{\alpha}} _m}{m} \, e^{- i \, m\, \sigma} \Big) \, \mathcal{E}_\nu^{m} ( z) \, .
\end{split}
\end{equation}
with  $z = t - \tau$, $\nu=2-2\,\alpha$, $1/2 \leq \nu < 1$, $\bm{\alpha} _m ^\ast = \bm{\alpha} _{-m}$ and $\widetilde{\bm{\alpha}} _m ^\ast = \widetilde{\bm{\alpha}} _{-m}$, $\forall m \in \mathbb{Z}$. Here,
	\begin{equation} \label{new-function}
\mathcal{E}_\nu^{m}(  z) := 
\sqrt{\frac{\pi\,|m|}{2}} \, z^\nu 
\left\{
\begin{aligned}
& H ^{(1)}_{-\nu} (|m| z) \, , \,\, m \in \mathbb{Z} ^- \, , \\
& H ^{(2)}_{-\nu} (|m| z) \, , \,\, m \in \mathbb{N} \, , \\
 \end{aligned}
\right.,
\end{equation}
which can be also written as
\be\label{New-functionBesselform}
\mathcal{E}_\nu^{m}( z) := 
\sqrt{\frac{\pi\,|m|}{2}} \, z^\nu 
\Big[J_{-\nu}(|m|z)-i\,\texttt{Sgn}(m)\,Y_{-\nu}(|m|z)
\Big]\,.
\ee
From Eq.\eqref{New-functionBesselform} one can see that the complex conjugate is $(\mathcal{E}_\nu^{m}( z) )^\ast=\mathcal{E}_\nu^{-m}( z) $. We already mention that the limit to recover the usual bosonic strings is given by taking $\alpha\rightarrow1$, (similarly $\nu\rightarrow1/2$) and $t\rightarrow \infty$. Therefore, It is useful to know
	\begin{equation} \label{new-functionlimit}
\mathcal{E}_{1/2}^{m}(  z) =e^{-imz}\quad,\quad\mathcal{E}_{-1/2}^{m}(  z) =\frac{i\,\texttt{Sgn}(m)}{z}\,e^{-imz} \,,\quad \forall\, m\in\mathbb{Z}-\{0\}.
\end{equation}
Here, $\texttt{Sgn} (m)$ is the sign function.

The embedding map can be also written in an asymptotic representation form 
\be
 \bm{X} (z, \sigma) = \bm{X}_0 -\sqrt{2\alpha'}\, \frac{ z^{2 \nu}}{2 \nu} \, \bm{\alpha} _0
  - i \,z^{\nu-1/2}\, \sqrt{\frac{\alpha'}{2}} \sum _{m \neq 0} e^{-i\,\delta_\nu(m)} \Big( \frac{\bm{\alpha} _m}{m} \, e^{i \, m\, \sigma} +  \frac{\widetilde{\bm{\alpha}} _m}{m} \, e^{- i \, m\, \sigma} \Big) \, e^{-imz}.\nonumber\\
\ee
when $z\gg |\nu^2-\frac{1}{4}|$. Here, we have used the asymptotic representation of the Hankel functions~\cite{Abram-Steg}
\be
H^{(1)}_\nu(z)\sim \sqrt{\frac{2}{\pi z}}\, e^{iz}\, e^{-i\frac{\pi}{2}(\nu-\frac{1}{2})}\quad\mbox{for $-\pi<\arg z<2\pi$}\,;\\
H^{(2)}_\nu(z)\sim \sqrt{\frac{2}{\pi z}}\, e^{-iz}\, e^{i\frac{\pi}{2}(\nu-\frac{1}{2})}\quad\mbox{for $-\pi<\arg z<2\pi$}\,,
\ee
which replaced in the function $\mathcal{E}_\nu^m(z)$ takes the form 
\be
\mathcal{E}_\nu^{m}( z)\sim z^{\nu-1/2}\,e^{-imz}\,e^{-i\,\delta_\nu(m)} \quad,\quad \mathcal{E}_{\nu-1}^{m}(z)\sim i\,\texttt{Sgn}(m)\,z^{\nu-3/2}\,e^{-imz}\,e^{-i\,\delta_\nu(m)}\,,
\ee
with $\delta_\nu(m)=\frac{\pi}{2}\texttt{Sgn}(m)(\nu-\frac{1}{2})$.

The corresponding derivatives of the embedding map $\bm{X}(\tau,\sigma)$ are given by
\be
\dot{\bm{X}} (z,\sigma)&= &\sqrt{2\alpha'}\, z^{2 \nu - 1} \, \bm{\alpha} _0\nonumber\\\label{Xdot}
&&- i \, z \, \sqrt{\frac{\alpha'}{2}} \sum _{m \neq 0} \texttt{Sgn} (m) \Big( \bm{\alpha} _m \, e^{i \, m\, \sigma} +  \widetilde{\bm{\alpha}} _m \, e^{- i \, m\, \sigma} \Big) \, \mathcal{E}_{\nu-1}^{m} ( z) \, ;\\ \label{Xprime}
\bm{X}' (z,\sigma)&=&  \, \sqrt{\frac{\alpha'}{2}} \sum _{m \neq 0}  \Big( \bm{\alpha} _m \, e^{i \, m\, \sigma} - \widetilde{\bm{\alpha}} _m \, e^{- i \, m\, \sigma} \Big) \, \mathcal{E}_\nu^{m} ( z)  \, .
\ee
Here, we have used $\dot{\bm{X}} = -\frac{\partial \bm{X}}{\partial z}$ and 
\be 
\frac{\partial}{\partial z}\mathcal{E}_\nu^{m} ( z) = - m\,\texttt{Sgn}(m) \, z \, \mathcal{E}_{\nu-1}^{m} ( z).
\ee

Also, we can define the following combinations
\be
E^\nu_m(z)&\equiv&\frac{z^{1/2-\nu}}{2}\left[ \mathcal{E}_{\nu}^{m} (z)+iz\,\texttt{Sgn}(m)\,\mathcal{E}_{\nu-1}^{m} (z) \right]\quad,\quad E^\nu_0(z)\equiv\frac{-z^{\nu-1/2}}{2}\,;\\
\widetilde{E}^\nu_m(z)&\equiv&\frac{z^{1/2-\nu}}{2}\left[ \mathcal{E}_{\nu}^{m} (z)-iz\,\texttt{Sgn}(m)\,\mathcal{E}_{\nu-1}^{m} (z) \right]\quad,\quad \widetilde{E}^\nu_0(z)\equiv\frac{z^{\nu-1/2}}{2}\,,
\ee
where in the limit $\nu\rightarrow1/2$, we have
\be
E^{1/2}_m(z)=0\quad&,&\quad E^{1/2}_0(z)=-\frac{1}{2}\,;\\
\widetilde{E}^{1/2}_m(z)=e^{-imz}\quad &,&\quad \widetilde{E}^{1/2}_0(z)=\frac{1}{2}\,,
\ee
and in the asymptotic limit are given by
\be
E^{\nu}_m(z)&\sim&0\quad ,\quad E^{\nu}_0(z)=\frac{-z^{\nu-1/2}}{2}\,;\\
\widetilde{E}^{\nu}_m(z)&\sim&e^{-imz}\,e^{-i\,\delta_\nu(m)}\quad ,\quad \widetilde{E}^{\nu}_0(z)=\frac{z^{\nu-1/2}}{2}\,.
\ee

Therefore, we can write
\be
\dot{\bm{X}}+\bm{X}'&=&2\,\partial_+\bm{X}=\sqrt{2\,\alpha'}\,z^{\nu-1/2}\sum_{m\in\mathbb{Z}}\Big(\widetilde{\bm{\alpha}}_m\widetilde{E}^\nu_m(z)\,e^{-im\sigma}-\bm{\alpha}_m\,E^\nu_m(z)\,e^{im\sigma}\Big)\,;\\
\dot{\bm{X}}-\bm{X}'&=&2\,\partial_-\bm{X}=\sqrt{2\,\alpha'}\,z^{\nu-1/2}\sum_{m\in\mathbb{Z}}\Big(\bm{\alpha}_m\widetilde{E}^\nu_m(z)\,e^{im\sigma}-\widetilde{\bm{\alpha}}_m\,E^\nu_m(z)\,e^{-im\sigma}\Big)\,.
\ee
Using the asymptotic approximation we can simplify the above equations as
\be
\dot{\bm{X}}+\bm{X}'&=&\sqrt{2\,\alpha'}\,z^{\nu-1/2}\sum_{m\in\mathbb{Z}}\ell_{m}^\nu(z)\,\widetilde{\bm{\alpha}}_m\,e^{-im(z+\sigma)}\,;\\
\dot{\bm{X}}-\bm{X}'&=&\sqrt{2\,\alpha'}\,z^{\nu-1/2}\sum_{m\in\mathbb{Z}}
\ell_{m}^\nu(z)\,\bm{\alpha}_m\,e^{-im(z-\sigma)}\,,
\ee
where
\be
\ell_{m}^\nu(z)=\exp\Big[-\frac{i\pi}{2}\left(\nu-\frac{1}{2}\right)\left(\texttt{Sgn}(m)+\frac{2i}{\pi}\,\delta_{0,m}\ln(z)\right)\Big].
\ee

Finally we can write $T_{\pm\pm}$ as
\be
T_{++}&=&\frac{z^{1-2\nu}}{4\alpha'\,\Gamma(2-2\nu)}\,||\dot{\bm{X}}+\bm{X}'||^2=\frac{z^{1-2\nu}}{4\alpha'\,\Gamma(2-2\nu)}\left(||\dot{\bm{X}}||^2+||\bm{X}'||^2+2\dot{\bm{X}}\cdot\bm{X}\right)\nonumber\\
&=&\frac{z^{1-2\nu}}{\alpha'\,\Gamma(2-2\nu)}\,\Big(\partial_+\,\bm{X}\cdot\partial_+\,\bm{X}\Big)\\
T_{--}&=&\frac{z^{1-2\nu}}{4\alpha'\,\Gamma(2-2\nu)}\,||\dot{\bm{X}}-\bm{X}'||^2=\frac{z^{1-2\nu}}{4\alpha'\,\Gamma(2-2\nu)}\left(||\dot{\bm{X}}||^2+||\bm{X}'||^2-2\dot{\bm{X}}\cdot\bm{X}\right)\nonumber\\
&=&\frac{z^{1-2\nu}}{\alpha'\,\Gamma(2-2\nu)}\,\Big(\partial_-\,\bm{X}\cdot\partial_-\,\bm{X}\Big)\,.
\ee
which in terms of the series expansion are given by
\be\label{T++full}
T_{++}&=&\frac{1}{2\,\Gamma(2-2\nu)}\sum_{n,m\in\mathbb{Z}}\Big(E_m^\nu(z)\,E_n^\nu(z)\,\bm{\alpha}_m\cdot\bm{\alpha}_n\,e^{i(m+n)\sigma}-E_m^\nu(z)\,\widetilde{E}_n^\nu(z)\,\bm{\alpha}_m\cdot\widetilde{\bm{\alpha}}_n\,e^{i(m-n)\sigma}\nonumber\\
&&-\,E_n^\nu(z)\,\widetilde{E}_m^\nu(z)\,\widetilde{\bm{\alpha}}_m\cdot\bm{\alpha}_m\,e^{-i(m-n)\sigma}+\widetilde{E}_m^\nu(z)\,\widetilde{E}_n^\nu(z)\,\widetilde{\bm{\alpha}}_m\cdot\widetilde{\bm{\alpha}}_n\,e^{-i(m+n)\sigma}\Big)\,;\\\label{T--full}
T_{--}&=&\frac{1}{2\,\Gamma(2-2\nu)}\sum_{n,m\in\mathbb{Z}}\Big(\widetilde{E}_m^\nu(z)\,\widetilde{E}_n^\nu(z)\,\bm{\alpha}_m\cdot\bm{\alpha}_n\,e^{i(m+n)\sigma}-E_m^\nu(z)\,\widetilde{E}_n^\nu(z)\,\widetilde{\bm{\alpha}}_m\cdot\bm{\alpha}_n\,e^{-i(m-n)\sigma}\nonumber\\
&&-\,E_n^\nu(z)\,\widetilde{E}_m^\nu(z)\,\bm{\alpha}_m\cdot\widetilde{\bm{\alpha}}_n\,e^{i(m-n)\sigma}+E_m^\nu(z)\,E_n^\nu(z)\,\widetilde{\bm{\alpha}}_m\cdot\widetilde{\bm{\alpha}}_n\,e^{-i(m+n)\sigma}\Big)\,,
\ee
where in the asymptotic limit can be written as
\be
T_{++}&=&\frac{1}{2\,\Gamma(2-2\nu)}\sum_{n,m\in\mathbb{Z}}\ell_{m}^\nu(z)\,\ell_n^\nu(z)\,\widetilde{\bm{\alpha}}_m\cdot\widetilde{\bm{\alpha}}_n\,e^{-i(m+n)(z+\sigma)}\,;\\
T_{--}&=&\frac{1}{2\,\Gamma(2-2\nu)}\sum_{n,m\in\mathbb{Z}}\ell_{m}^\nu(z)\,\ell_n^\nu(z)\,\bm{\alpha}_m\cdot\bm{\alpha}_n\,e^{-i(m+n)(z-\sigma)}\,.
\ee
An important crossed check of our solution is to verify whether the energy momentum tensor in Eq. \eqref{T++full} and \eqref{T--full} in fact satisfy the homogeneous equation in \eqref{conservationlaw}. This is indeed verified in the Appendix~\ref{Appen:ContiEqu}. The explicit forms of the energy momentum tensor will be used in the next section to define the Fractional Virasoro Operators.

\section{Hamiltonian, Fractional Virasoro, and $\sigma$-translation operators}\label{Sec2}
In this section we want to review the computations of the Hamiltonian and the operator making rigid translations along the $\sigma$-direction. Also we want to introduce the Fractional Virasoro Operators.

The Hamiltonian is defined by
\be
H(z)=\frac{z^{1-2\nu}}{4\pi\alpha'\,\Gamma(2-2\nu)}\int_{0}^{2\pi}d\sigma \left(||\dot{\bm{X}}||^2+||\bm{X}'||^{2}\right),
\ee
which is equal to [ref]
\be
H(z)=\frac{1}{2\pi}\,\int_0^{2\pi}d\sigma\, \left(T_{++}+T_{--}\right)\equiv \left[L_0(\nu;z)+\tilde{L}_0(\nu;z)\right].
\ee
Here, we have
\begin{eqnarray}\label{L0no}
L_0(\nu;z)\equiv\frac{1}{4} \sum _{m \in\mathbb{Z}}\left[ \bm{\alpha} _m\cdot \bm{\alpha} _{-m}\,\mathcal{G}_{\nu1}^m(z)+\bm{\alpha} _m\cdot \widetilde{\bm{\alpha}} _{m}\,\mathcal{G}_{\nu2}^m(z)\right],\\\label{L0tildeno}
\widetilde{L}_0(\nu;z)\equiv\frac{1}{4} \sum _{m\in\mathbb{Z}}\left[ \widetilde{\bm{\alpha}}_m\cdot \widetilde{\bm{\alpha}}_{-m}\,\mathcal{G}_{\nu1}^m(z)+\widetilde{\bm{\alpha}} _m\cdot \bm{\alpha} _{m}\,\mathcal{G}_{\nu2}^m(z)\right]\,,
\end{eqnarray} 
where we have used the fact that $\bm{\alpha}_0=\widetilde{\bm{\alpha}}_0$, and we have also defined $\mathcal{G}^0_{\nu1}(z)=\frac{2z^{2\nu-1}}{\Gamma(2-2\nu)}$, $\mathcal{G}_{\nu2}^0(z)=0$, and for $m\neq 0$
\be
\mathcal{G}_{\nu1}^m(z)&\equiv&\frac{z^{1-2\nu}}{\Gamma(2-2\nu)}\left[ z^2\,\mathcal{E}_{\nu - 1}^m(z) \,\mathcal{E}_{\nu - 1}^{-m}( z)+\mathcal{E}_{\nu}^m(z) \,\mathcal{E}_\nu^{-m}( z) \right]\,,\nonumber\\[3mm]
\mathcal{G}_{\nu2}^m(z)&\equiv&\frac{-z^{1-2\nu}}{\Gamma(2-2\nu)}\left[ z^2\,\mathcal{E}_{\nu - 1}^m (z) \,\mathcal{E}_{\nu - 1}^m( z)+\mathcal{E}_{\nu}^m (z) \,\mathcal{E}_{\nu}^m(z)\right]\, .\nonumber
\ee
Although this definition of the Virasoro operators of order zero seems natural is not unique. We can always make the redefinitions
\be\label{transf1}
L_{0}(\nu;z)&\rightarrow& L_0(\nu;z)+a(z;\nu)\\[2mm]\label{transf2}
\widetilde{L}_0(\nu;z)&\rightarrow&\widetilde{L}_0(\nu;z)-a(z;\nu)\,,
\ee
and $H(z)$ remain invariant. In order to uni-vocally define the $L_{0}$ $(\widetilde{L}_{0})$, we not only need to look at the Hamiltonian, we also need to take into consideration the operator making rigid translation along the $\sigma$-direction, \ie the operator $P_\sigma$ satisfying
\be\label{Translationsigma}
\left\lbrace P_{\sigma}\,,\, X^\mu(\tau,\sigma)\right\rbrace_{P.B.}=-\partial_\sigma X^\mu(\tau,\sigma),
\ee
where $\{\cdot\,,\,\cdot\}_{P.B.}$ is the Poisson bracket. This is due to the fact that the constraints \eqref{our-EOM-sym-2} in the factional conformal gauge, form a closed algebra under the Poisson brackets but unlike the case of bosonic string theory, the Virasoro operators are not constant in $\tau$ (or equivalently in $z$). Taking this into consideration, we proceed to find $P_\sigma$. Using the canonical Poisson Bracket 
\be
\left\lbrace X^\mu(z,\sigma),\mathcal{P}^{\tau\nu}(z,\sigma')\right\rbrace_{P.B.}=\eta^{\mu\nu}\,\delta(\sigma-\sigma')\,,
\ee
with $\mathcal{P}^{\tau\nu}(z,\sigma')=\frac{z^(1-2\nu)}{2\pi\alpha'\,\Gamma(2-2\nu)}\,\dot{\bm{X}}(z,\sigma')$, 
it is easy to check that the operator generating the transformation \eqref{Translationsigma} is given by
\be
P_{\sigma}&=&\int_{0}^{2\pi}d\sigma \,\bm{\mathcal{P}}^\tau(z,\sigma) \cdot\bm{X}'(z,\sigma).
\ee

It is well known in bosonic string theory that the operator generating this translation is exactly the same and its Fourier decomposition is given by 
\be
P_\sigma=L_0-\widetilde{L}_0=\frac{1}{2}\sum_{n\in\mathbb{Z}}\Big(\bm{\alpha}_{-n}\cdot\bm{\alpha}_n-\bm{\widetilde{\alpha}}_{-n}\cdot\bm{\widetilde{\alpha}}_n\Big).
\ee
In our theory, we have not broken the reparametrisation invariance along the $\sigma$-direction, therefore we still have the conserved space-time momentum. Then, we might expect that as in the usual bosonic string theory $P_\sigma$ takes a similar form and do not depend on $z$. Making the natural definition of the Virasoro operators of zero order in Eqs. \eqref{L0no} and \eqref{L0tildeno} a wrong definition. We then introduce the proper \textit{\textbf{Fractional Virasoro Operators of order $p$}} as
\begin{definition}{\textbf{Fractional Virasoro Operators of order $p$}.} The fractional Virasoro operators are defined by
\be
\widetilde{L}_{p,\nu}(z)&\equiv&\frac{1}{2\pi}\int_0^{2\pi} d\sigma\, e^{ip\sigma}\,T_{++}(z,\sigma)\,;\\
L_{p,\nu}(z)&\equiv&\frac{1}{2\pi}\int_0^{2\pi} d\sigma\, e^{-ip\sigma}\,T_{--}(z,\sigma)\,,
\ee
where using the expansions \eqref{T++full} and \eqref{T--full} can be written as
\be\label{VirasoroLtilde}
\widetilde{L}_{p,\nu}(z)&=&\frac{1}{2\,\Gamma(2-2\nu)}\sum_{n\in\mathbb{Z}}\Big(E_{-n-p}^\nu\,E_n^\nu\,\bm{\alpha}_{-n-p}\cdot\bm{\alpha}_n-E_{n-p}^\nu\,\widetilde{E}_n^\nu\,\bm{\alpha}_{n-p}\cdot\widetilde{\bm{\alpha}}_n-E_{n}^\nu\,\widetilde{E}_{n+p}^\nu\,\widetilde{\bm{\alpha}}_{n+p}\cdot\bm{\alpha}_{n}\nonumber\\
&&+\,\widetilde{E}_{p-n}^\nu\,\widetilde{E}_n^\nu\,\widetilde{\bm{\alpha}}_{p-n}\cdot\widetilde{\bm{\alpha}}_n\Big)\,;\\\label{VirasoroL}
L_{p,\nu}(z)&=&\frac{1}{2\,\Gamma(2-2\nu)}\sum_{n\in\mathbb{Z}}\Big(\widetilde{E}_{p-n}^\nu\,\widetilde{E}_n^\nu\,\bm{\alpha}_{p-n}\cdot\bm{\alpha}_n-E_{n-p}^\nu\,\widetilde{E}_n^\nu\,\widetilde{\bm{\alpha}}_{n-p}\cdot\bm{\alpha}_n-\,E_{n}^\nu\,\widetilde{E}_{n+p}^\nu\,\bm{\alpha}_{n+p}\cdot\widetilde{\bm{\alpha}}_{n}\nonumber\\
&&+\,E_{-p-n}^\nu\,E_n^\nu\,\widetilde{\bm{\alpha}}_{-p-n}\cdot\widetilde{\bm{\alpha}}_n\Big)\,.
\ee
\end{definition}
As we clearly see these operators depend explicitly on $z$. Also notice that $\left[L_{p,\nu}(z)\right]^\ast=L_{-p,\nu}(z)$ and $\left[\widetilde{L}_{p,\nu}(z)\right]^\ast=\widetilde{L}_{-p,\nu}(z)$ keeping the Hermiticity of the theory still valid. 

In order to verify that whether this definition is the correct one or not, we need to compute $P_\sigma$ and the Hamiltonian $H(z)$. For $p=0$, we have
\be
\widetilde{L}_{0,\nu}(z)&=&\frac{1}{2}\left[L_0(\nu;z)+\widetilde{L}_0(\nu;z)\right]-\sum_{n\neq0}\mathcal{G}_{0\nu}^n(z)\left[\bm{\alpha}_{-n}\cdot\bm{\alpha}_n-\widetilde{\bm{\alpha}}_{-n}\cdot\widetilde{\bm{\alpha}}_n\right]\,;\\
L_{0,\nu}(z)&=&\frac{1}{2}\left[L_0(\nu;z)+\widetilde{L}_0(\nu;z)\right]+\sum_{n\neq0}\mathcal{G}_{0\nu}^n(z)\left[\bm{\alpha}_{-n}\cdot\bm{\alpha}_n-\widetilde{\bm{\alpha}}_{-n}\cdot\widetilde{\bm{\alpha}}_n\right],
\ee
with
$$\mathcal{G}_{0\nu}^n(z)=\frac{i\,z^{2-2\nu}\,\text{Sgn}(n)}{8\,\Gamma(2-2\nu)}\Big(\mathcal{E}_\nu^n(z)\,\mathcal{E}_{\nu-1}^{-n}(z)-\mathcal{E}_\nu^{-n}(z)\,\mathcal{E}_{\nu-1}^{n}(z)\Big).$$
In fact, using the relation \eqref{Erelation} in the Appendix \ref{Appen:Prop} we have
\be
\widetilde{L}_{0,\nu}(z)&=&\frac{1}{2}\left[L_0(\nu;z)+\widetilde{L}_0(\nu;z)\right]-\frac{1}{4\,\Gamma(2-2\nu)}\sum_{n\neq0}\Big(\bm{\alpha}_{-n}\cdot\bm{\alpha}_n-\widetilde{\bm{\alpha}}_{-n}\cdot\widetilde{\bm{\alpha}}_n\Big)\,;\\[2mm]
L_{0,\nu}(z)&=&\frac{1}{2}\left[L_0(\nu;z)+\widetilde{L}_0(\nu;z)\right]+\frac{1}{4\,\Gamma(2-2\nu)}\sum_{n\neq0}\Big(\bm{\alpha}_{-n}\cdot\bm{\alpha}_n-\widetilde{\bm{\alpha}}_{-n}\cdot\widetilde{\bm{\alpha}}_n\Big)\,.
\ee
Therefore, we see that the Hamiltonian is given by
\be
H(z)&=&\left[L_0(\nu;z)+\widetilde{L}_0(\nu;z)\right]=\left[L_{0,\nu}(z)+\widetilde{L}_{0,\nu}(z)\right]\,.
\ee
In this new representation of the Virasoro operators the redefinitions \eqref{transf1} and \eqref{transf2} are still valid  but the operator generating this transformation is given by
\be
P_{\sigma}&=&\int_{0}^{2\pi}d\sigma \,\bm{\mathcal{P}}^\tau(z,\sigma) \cdot\bm{X}'(z,\sigma)\nonumber\\
&=&\frac{z^{1-2\nu}}{4\pi\alpha'\,\Gamma(2-2\nu)}\int_0^{2\pi}d\sigma\left(2\,\dot{\bm{X}}\cdot\bm{X}'\right)\nonumber\\
&=&\frac{1}{2\pi}\,\int_0^{2\pi}d\sigma\, \left(T_{++}-T_{--}\right).
\ee
Using the definition of $L_{0,\nu}$ and $\widetilde{L}_{0,\nu}$,  $P_\sigma$ can be written as
\be
P_\sigma= L_{0,\nu}(z)-\widetilde{L}_{0,\nu}(z)=\frac{1}{2\,\Gamma(2-2\nu)}\,\sum_{n\in\mathbb{Z}}\Big(\bm{\alpha}_{-n}\cdot\bm{\alpha}_n-\widetilde{\bm{\alpha}}_{-n}\cdot\widetilde{\bm{\alpha}}_n\Big)\,,
\ee
where we have used the fact $\bm{\alpha}_0=\widetilde{\bm{\alpha}}_0$. Therefore with this new definition we clearly see that $P_\sigma$ takes exactly the same form, up to normalization factor for oscillator modes, of the usual bosonic string theory as expected. Proving that this definition is the correct one. It is worth notice that $\widetilde{L}_{p,\nu}(z)$ and $L_{p,\nu}(z)$ are independent. This fact is not completely clear from Eqs.~\eqref{VirasoroLtilde} and~\eqref{VirasoroL} but it will be explained in Sec.\ref{Sec4}.

\section{The $\alpha(\widetilde{\alpha})$-oscillator algebra}\label{Sec3}
In this section we focus on the computation of the algebra satified by the $\alpha$ and $\widetilde{\alpha}$ oscillators. 

We know the canonical  Poisson brackets are
\be\label{brack1}
\left\lbrace X^\mu(z,\sigma),X^{\nu}(z,\sigma')\right\rbrace_{P.B.}=0\quad;\quad\left\lbrace \mathcal{P}^{\tau\mu}(z,\sigma),\mathcal{P}^{\tau\nu}(z,\sigma')\right\rbrace_{P.B.}=0\,;
\ee
\be\label{brack2}
\left\lbrace X^\mu(z,\sigma),\mathcal{P}^{\tau\nu}(z,\sigma')\right\rbrace_{P.B.}=\eta^{\mu\nu}\,\delta(\sigma-\sigma')\,,
\ee
where the canonical conjugate momentum is given by
\be 
\bm{\mathcal{P}}^\tau (z , \sigma) = \frac{z^{ 1-2\nu}}{2 \pi \alpha' \, \Gamma (2-2\nu)} \, \dot{\bm{X}} (z, \sigma) \, .\nonumber
\ee 

As we mentioned in Section \ref{Sec2} the total space-time momentum 
\be 
\bm{p}=\int_0^{2\pi}d\sigma\,\bm{\mathcal{P}}^\tau (z , \sigma) =\sqrt{\frac{2}{\alpha'}} \frac{\bm{\alpha}_0}{\Gamma (2-2\nu)} \, ,
\ee
is conserved, as we clearly from the equation above. Using the momentum $\bm{p}$, the embedding map $\bm{X}(z,\sigma)$ and the canonical conjugate momentum  $\bm{\mathcal{P}}^\tau (z , \sigma) $ can be written as
\begin{eqnarray} 
\begin{split}
 \bm{X} (z, \sigma) &=& \bm{X}_0 - \frac{\alpha'\,z^{2\nu}\,\Gamma(2-2\nu)}{2 \nu} \, \bm{p}
  - i \, \sqrt{\frac{\alpha'}{2}} \sum _{m \neq 0}  \Big( \frac{\bm{\alpha} _m}{m} \, e^{i \, m\, \sigma} +  \frac{\widetilde{\bm{\alpha}} _m}{m} \, e^{- i \, m\, \sigma} \Big) \, \mathcal{E}_\nu^{m} ( z) \,;\nonumber\\
\bm{\mathcal{P}}^\tau (z , \sigma)&=&\frac{\bm{p}}{2\pi}- \frac{iz^{2-2\nu}}{2\pi\,\sqrt{2\alpha'}\,\Gamma(2-2\nu)}  \sum _{n \neq 0} \texttt{Sgn} (n) \Big( \bm{\alpha} _n \, e^{i \,n\, \sigma} +  \widetilde{\bm{\alpha}} _n \, e^{- i \, n\, \sigma} \Big) \, \mathcal{E}_{\nu-1}^{n} ( z)\,.
\end{split}
\end{eqnarray}
Thus, replacing these expansions in the brackets and using the Dirac delta representation
$$\delta(\sigma-\sigma')=\frac{1}{2\pi}\sum_{m\in\mathbb{Z}}e^{im(\sigma-\sigma')}\,,$$
we find the following commutation relations:

\subsubsection{Natural commutators}
From the brackets in Eq.~\eqref{brack1}, we find the relations
\be
\left\lbrace X^\mu_0 , \widetilde{\alpha}^\rho _n \right\rbrace_{P.B.}=\left\lbrace X^\mu_0 , \alpha^\rho _n \right\rbrace_{P.B.}=0\quad;\,\quad\left\lbrace p^\mu , \widetilde{\alpha}^\rho _n \right\rbrace_{P.B.}=\left\lbrace p^\mu, \alpha^\rho _n \right\rbrace_{P.B.}=0\,.
\ee
Together with the bracket
\be
\left\lbrace\alpha_m^\mu , \widetilde{\alpha}^\rho _n \right\rbrace_{P.B.}=0\,\,, \quad\forall \,m,n\in\mathbb{Z}-\{0\}\,,
\ee
re-confirming that the $\alpha$-oscillators are independent.

\subsubsection{$\alpha$'s commutators}
From the canonical Poisson bracket~\eqref{brack2}, we have several cases:
\begin{itemize}
\item Case for order $m=0$:
\be
\left\lbrace X^\mu_0 - \frac{\alpha'\,z\,\Gamma(2-2\nu)}{2 \nu} \,p^\mu, p^\rho\right\rbrace_{P.B.}=\eta^{\mu\rho}\,,
\ee
where we can deduce the relation
\be
\left\lbrace X^\mu_0 , p^\rho\right\rbrace_{P.B.}=\eta^{\mu\rho}\,.
\ee
Thus, we find that $X_0^\mu$ and $p^\mu$, the center of mass position and the momentum, are canonically conjugate, as expected.
\item In the case with only one sum we recover the same relations as in the natural commutators case. Instead for the terms with double sums, we have
\be
&&-\frac{z^{2-2\nu}}{2\Gamma(2-2\nu)}\sum_{m,n\neq0}\frac{\texttt{Sgn} (n) }{m}\mathcal{E}_\nu^{m} \, \mathcal{E}_{\nu-1}^{n}\,\left(\left\lbrace\alpha^\mu _m,\alpha^\rho _n   \right\rbrace_{P.B.}\,e^{i \, (m\, \sigma+n\,\sigma')}+\right.\nonumber\\
&&\left.+\,\left\lbrace\alpha^\mu _m,\widetilde{\alpha}^\rho _n    \right\rbrace_{P.B.}\, e^{i \, (m\, \sigma-n\,\sigma')}+ \left\lbrace\widetilde{\alpha}^\mu _m,\alpha^\rho _n   \right\rbrace_{P.B.}\, e^{-i \, (m\, \sigma-n\,\sigma')}+\left\lbrace\widetilde{\alpha}^\mu _m,\widetilde{\alpha}^\rho _n  \right\rbrace_{P.B.}\, e^{-i \, (m\, \sigma+n\,\sigma')}\right)=\nonumber\\
&&  =\,\eta^{\mu\rho}\sum_{m\neq0}e^{im(\sigma-\sigma')}\,.\nonumber
\ee
We know that the oscillators are independent, therefore
\be
&&-\frac{z^{2-2\nu}}{2\Gamma(2-2\nu)}\sum_{m,n\neq0}\frac{\texttt{Sgn} (n) }{m}\mathcal{E}_\nu^{m} \, \mathcal{E}_{\nu-1}^{n}\,\left(\left\lbrace\alpha^\mu _m,\alpha^\rho _n   \right\rbrace_{P.B.}\,e^{i \, (m\, \sigma+n\,\sigma')}+\left\lbrace\widetilde{\alpha}^\mu _m,\widetilde{\alpha}^\rho _n  \right\rbrace_{P.B.}\, e^{-i \, (m\, \sigma+n\,\sigma')}\right) =\nonumber\\
&&=\,\eta^{\mu\rho}\sum_{m\neq0}e^{im(\sigma-\sigma')}\,.\nonumber
\ee
This relation is different from zero for $m+n=0$, finally writing
\be\label{alphacom}
&&\frac{z^{2-2\nu}}{2\Gamma(2-2\nu)}\sum_{m\neq0}\frac{\texttt{Sgn} (m) }{m}\left(\mathcal{E}_\nu^{m} \, \mathcal{E}_{\nu-1}^{-m}\,\left\lbrace\alpha^\mu _m,\alpha^\rho _{-m}   \right\rbrace_{P.B.}+\,\mathcal{E}_\nu^{-m} \, \mathcal{E}_{\nu-1}^{m}\,\left\lbrace\widetilde{\alpha}^\mu _{-m},\widetilde{\alpha}^\rho _{m} \right\rbrace_{P.B.}\right)e^{i \, m( \sigma-\sigma')}=\nonumber\\
&&=\eta^{\mu\rho}\sum_{m\neq0}e^{im(\sigma-\sigma')}\,.
\ee
If we set 
\be
\left\lbrace\alpha^\mu _m,\alpha^\rho _{-m}   \right\rbrace_{P.B.}=i\,m\,\Gamma(2-2\nu)\,\eta^{\mu\rho}\quad,\quad\left\lbrace\widetilde{\alpha}^\mu _{-m},\widetilde{\alpha}^\rho _{m} \right\rbrace_{P.B.}=-i\,m\,\Gamma(2-2\nu)\,\eta^{\mu\rho}\,,\nonumber
\ee
together with the relation \eqref{Erelation}, the equation \eqref{alphacom} is satisfy. Finally, we can write
\be\label{alphacommfinal}
&\left\lbrace\alpha^\mu _m,\alpha^\rho _{n}   \right\rbrace_{P.B.}=i\,m\,\delta_{m+n,0}\,\Gamma(2-2\nu)\,\eta^{\mu\rho}\,\,,\,\,\left\lbrace\widetilde{\alpha}^\mu _{m},\widetilde{\alpha}^\rho _{n} \right\rbrace_{P.B.}=i\,m\,\delta_{m+n,0}\,\Gamma(2-2\nu)\,\eta^{\mu\rho}&\nonumber\\[2mm]
&\bm{\forall \,n,m\in\mathbb{Z}}\,.&\nonumber
\ee
\end{itemize}
This proves that up to a renormalization factor the modes $\alpha$ and $\widetilde{\alpha}$ satisfy two independent harmonic oscillator algebra. In the next section we will make use of this relations for computing the algebra that the Fractional Virasoro Operators satisfy.

\section{Fractional Virasoro algebra}\label{Sec4}

In this section we focus on the computation of the Poisson Brackets of the Fractional Virasoro Operators defined in \eqref{VirasoroLtilde} and \eqref{VirasoroL}. It is useful to know first the following commutators
\be\label{comm1}
\left\lbrace L_{p,\nu}(z),\alpha^\mu _{m}   \right\rbrace_{P.B.}&=&-im\Big[ \widetilde{E}_{p+m}^\nu\,\widetilde{E}_{-m}^\nu\,\alpha^\mu _{p+m}- E_{-m-p}^\nu\,\widetilde{E}_{-m}^\nu\,\widetilde{\alpha}^\mu _{-p-m}\Big]\,;\\\label{comm2}
\left\lbrace L_{p,\nu}(z),\widetilde{\alpha}^\mu _{m}   \right\rbrace_{P.B.}&=&-im\Big[ E_{m-p}^\nu\,E_{-m}^\nu\,\widetilde{\alpha}^\mu _{m-p}- E_{-m}^\nu\,\widetilde{E}_{p-m}^\nu\,\alpha^\mu _{p-m}\Big]\,;\\\label{comm3}
\left\lbrace \widetilde{L}_{p,\nu}(z),\widetilde{\alpha}^\mu _{m}   \right\rbrace_{P.B.}&=& -im\Big[ \widetilde{E}_{p+m}^\nu\,\widetilde{E}_{-m}^\nu\,\widetilde{\alpha}^\mu _{p+m}- E_{-m-p}^\nu\,\widetilde{E}_{-m}^\nu\,\alpha^\mu _{-p-m}\Big]\,;\\\label{comm4}
\left\lbrace\widetilde{ L}_{p,\nu}(z),\alpha^\mu _{m}   \right\rbrace_{P.B.}&=&-im\Big[ E_{m-p}^\nu\,E_{-m}^\nu\,\alpha^\mu _{m-p}- E_{-m}^\nu\,\widetilde{E}_{p-m}^\nu\,\widetilde{\alpha}^\mu _{p-m}\Big]\,.
\ee
The proof of these relations is as follows:
\begin{proof}
\be
\left\lbrace L_{p,\nu}(z),\alpha^\mu _{m}   \right\rbrace_{P.B.}&=&\frac{1}{2\,\Gamma(2-2\nu)}\sum_{n\in\mathbb{Z}}\Big(\widetilde{E}_{p-n}^\nu\,\widetilde{E}_n^\nu\,\left\lbrace \bm{\alpha}_{p-n}\cdot\bm{\alpha}_n,\alpha^\mu _{m}   \right\rbrace_{P.B.}-E_{n-p}^\nu\,\widetilde{E}_n^\nu\,\left\lbrace \widetilde{\bm{\alpha}}_{n-p}\cdot\bm{\alpha}_n,\alpha^\mu _{m}   \right\rbrace_{P.B.}\nonumber\\
&&-\,E_{n}^\nu\,\widetilde{E}_{n+p}^\nu\,\left\lbrace\bm{\alpha}_{n+p}\cdot\widetilde{\bm{\alpha}}_{n},\alpha^\mu _{m}\right\rbrace_{P.B.}+E_{-p-n}^\nu\,E_n^\nu\,\left\lbrace\widetilde{\bm{\alpha}}_{-p-n}\cdot\widetilde{\bm{\alpha}}_n,\alpha^\mu _{m}   \right\rbrace_{P.B.}\Big)\nonumber\\
&=&\frac{1}{2\,\Gamma(2-2\nu)}\sum_{n\in\mathbb{Z}}\Big(\widetilde{E}_{p-n}^\nu\,\widetilde{E}_n^\nu\,\left\lbrace \bm{\alpha}_{p-n}\cdot\bm{\alpha}_n,\alpha^\mu _{m}   \right\rbrace_{P.B.}-E_{n-p}^\nu\,\widetilde{E}_n^\nu\,\widetilde{\alpha}^\rho_{n-p}\left\lbrace \alpha^\rho_n,\alpha^\mu _{m}   \right\rbrace_{P.B.}\nonumber\\
&&-\,E_{n}^\nu\,\widetilde{E}_{n+p}^\nu\,\left\lbrace \alpha_{n+p}^\rho,\alpha^\mu _{m}\right\rbrace_{P.B.}\,\widetilde{\alpha}^\rho_{n}\Big)\,.
\ee
Here, we have used the independence of the oscillators $\alpha$ and $\widetilde{\alpha}$. Replacing the commutators \eqref{alphacommfinal}, we find
\be
\left\lbrace L_{p,\nu}(z),\alpha^\mu _{m}   \right\rbrace_{P.B.}&=&-im\Big[ \widetilde{E}_{p+m}^\nu\,\widetilde{E}_{-m}^\nu\,\alpha^\mu _{p+m}- E_{-m-p}^\nu\,\widetilde{E}_{-m}^\nu\,\widetilde{\alpha}^\mu _{-p-m}\Big]\,.
\ee
In a similar manner, we compute
\be
\left\lbrace L_{p,\nu}(z),\widetilde{\alpha}^\mu _{m}   \right\rbrace_{P.B.}&=&\frac{1}{2\,\Gamma(2-2\nu)}\sum_{n\in\mathbb{Z}}\Big(\widetilde{E}_{p-n}^\nu\,\widetilde{E}_n^\nu\,\left\lbrace \bm{\alpha}_{p-n}\cdot\bm{\alpha}_n,\widetilde{\alpha}^\mu _{m}   \right\rbrace_{P.B.}-E_{n-p}^\nu\,\widetilde{E}_n^\nu\,\left\lbrace \widetilde{\bm{\alpha}}_{n-p}\cdot\bm{\alpha}_n,\widetilde{\alpha}^\mu _{m}   \right\rbrace_{P.B.}\nonumber\\
&&-\,E_{n}^\nu\,\widetilde{E}_{n+p}^\nu\,\left\lbrace\bm{\alpha}_{n+p}\cdot\widetilde{\bm{\alpha}}_{n},\widetilde{\alpha}^\mu _{m}\right\rbrace_{P.B.}+E_{-p-n}^\nu\,E_n^\nu\,\left\lbrace\widetilde{\bm{\alpha}}_{-p-n}\cdot\widetilde{\bm{\alpha}}_n,\widetilde{\alpha}^\mu _{m}   \right\rbrace_{P.B.}\Big)\nonumber\\
&=&\frac{1}{2\,\Gamma(2-2\nu)}\sum_{n\in\mathbb{Z}}\Big(E_{-p-n}^\nu\,E_n^\nu\,\left\lbrace\widetilde{\bm{\alpha}}_{-p-n}\cdot\widetilde{\bm{\alpha}}_n,\widetilde{\alpha}^\mu _{m}   \right\rbrace_{P.B.}-E_{n-p}^\nu\,\widetilde{E}_n^\nu\,\left\lbrace \widetilde{\alpha}^\rho_{n-p},\widetilde{\alpha}^\mu _{m}   \right\rbrace_{P.B.}\alpha_n^\rho\nonumber\\
&&-\,E_{n}^\nu\,\widetilde{E}_{n+p}^\nu\,\alpha_{n+p}^\rho\,\left\lbrace\widetilde{\alpha}^\rho_{n},\widetilde{\alpha}^\mu _{m}\right\rbrace_{P.B.}\Big)\nonumber\\
&=&-im\Big[ E_{m-p}^\nu\,E_{-m}^\nu\,\widetilde{\alpha}^\mu _{m-p}- E_{-m}^\nu\,\widetilde{E}_{p-m}^\nu\,\alpha^\mu _{p-m}\Big]\,.
\ee
All this computation can be repeated for the $\widetilde{L}$ operators.
\end{proof}

The next step is to use the commutators \eqref{comm1}-\eqref{comm4} to compute the algebra satisfied by the Virasoro Operators:
\be\label{ViraAlgebraL}
\left\lbrace L_{p,\nu}(z),L_{q,\nu}(z)  \right\rbrace_{P.B.}&=&\frac{1}{2\,\Gamma(2-2\nu)}\sum_{n\in\mathbb{Z}}\Big(\widetilde{E}_{q-n}^\nu\,\widetilde{E}_n^\nu\,\left\lbrace L_{p,\nu}(z),\bm{\alpha}_{q-n}\cdot\bm{\alpha}_n\right\rbrace_{P.B.}-\nonumber\\
&&-\,E_{n-q}^\nu\,\widetilde{E}_{n}^\nu\,\left\lbrace L_{p,\nu}(z),\widetilde{\bm{\alpha}}_{n-q}\cdot\bm{\alpha}_n\right\rbrace_{P.B.}-E_{n}^\nu\,\widetilde{E}_{n+q}^\nu\,\left\lbrace L_{p,\nu}(z),\bm{\alpha}_{n+q}\cdot\widetilde{\bm{\alpha}}_n\right\rbrace_{P.B.}\nonumber\\
&&+\,E_{n}^\nu\,E_{-n-q}^\nu\,\left\lbrace L_{p,\nu}(z),\widetilde{\bm{\alpha}}_{-n-q}\cdot\widetilde{\bm{\alpha}}_n\right\rbrace_{P.B.}\Big)\,.
\ee
Using the relations \eqref{comm1}-\eqref{comm4}, we arrived
\be
\left\lbrace L_{p,\nu}(z),L_{q,\nu}(z)  \right\rbrace_{P.B.}&=&\frac{-i}{2\,\Gamma(2-2\nu)}\sum_{k\in\mathbb{Z}}\left\lbrace K_{k-p\,,\,k-q}^\nu(z)\,\widetilde{E}_{p+q-k}^\nu\,\widetilde{E}_{k}^\nu\,\bm{\alpha} _{p+q-k}\cdot\bm{\alpha}_{k}-\right.\nonumber\\
&&-\,K^\nu_{k-p\,,\,k-q}(z)\,E_{k-p-q}^\nu\,\widetilde{E}_{k}^\nu\,\widetilde{\bm{\alpha}}_{k-p-q}\cdot\bm{\alpha}_{k}-\nonumber\\
&&-\,K^\nu_{-k-p\,,\,-k-q}(z)\,\widetilde{E}_{k+p+q}^\nu\,E_{k}^\nu\,\bm{\alpha}_{k+p+q}\cdot\widetilde{\bm{\alpha}}_{k}+\nonumber\\\label{almostVira}
&&\left.+\,K^\nu_{-k-p\,,\,-k-q}(z)\,E_{-k-p-q}^\nu\,E_{k}^\nu\,\widetilde{\bm{\alpha}}_{-k-p-q}\cdot\widetilde{\bm{\alpha}}_{k}\right\rbrace\,.
\ee
The explicit computation of this relations is given in Appendix~\ref{Appen:ViraSums}. Here, we have defined
\be
K_{m,n}^\nu(z)\equiv\Big[m\,\left(\widetilde{E}_{-m}^\nu\,\widetilde{E}_{m}^\nu-E_{-m}^\nu\,E_{m}^\nu\right)-n\,\left(\widetilde{E}_{-n}^\nu\,\widetilde{E}_{n}^\nu-E_{-n}^\nu\,E_{n}^\nu\right)\Big]\,.
\ee
Thus, using the relations \eqref{Erelation}, \eqref{EE}, and \eqref{X} in the Appendix \ref{Appen:Prop}, we find 
\be
\widetilde{E}_{-m}^\nu\,\widetilde{E}_{m}^\nu-E_{-m}^\nu\,E_{m}^\nu=1\quad\forall\, m \neq0\,,
\ee
implying  that $K^\nu_{m,n}(z)=(m-n)$ for any $m,n\in\mathbb{Z}$. Therefore Eq. \eqref{almostVira} turns out to be
\be
\left\lbrace L_{p,\nu}(z),L_{q,\nu}(z)   \right\rbrace_{P.B.}&=&i(p-q)\,L_{p+q,\nu}(z)\nonumber\,.
\ee
We can repeat the computation for $\widetilde{L}$ operators
\be
\left\lbrace \widetilde{L}_{p,\nu}(z),\widetilde{L}_{q,\nu}(z)  \right\rbrace_{P.B.}&=&\frac{-i}{2\,\Gamma(2-2\nu)}\sum_{k\in\mathbb{Z}}\left\lbrace K_{k+q\,,\,k+p}^\nu(z)\,E_{-k-p-q}^\nu\,E_{k}^\nu\,\bm{\alpha} _{-k-p-q}\cdot\bm{\alpha}_{k}-\right.\nonumber\\
&&-\,K^\nu_{k+q\,,\,k+p}(z)\,\widetilde{E}_{k+p+q}^\nu\,E_{k}^\nu\,\widetilde{\bm{\alpha}}_{k+p+q}\cdot\bm{\alpha}_{k}-\nonumber\\
&&-\,K^\nu_{k-p\,,\,k-q}(z)\,E_{k-p-q}^\nu\,\widetilde{E}_{k}^\nu\,\bm{\alpha}_{k-p-q}\cdot\widetilde{\bm{\alpha}}_{k}+\nonumber\\\label{almostVira}
&&\left.+\,K^\nu_{k-p\,,\,k-q}(z)\,\widetilde{E}_{p+q-k}^\nu\,\widetilde{E}_{k}^\nu\,\widetilde{\bm{\alpha}}_{p+q-k}\cdot\widetilde{\bm{\alpha}}_{k}\right\rbrace\,,
\ee
where using the same relation as before we can write
\be
\left\lbrace \widetilde{L}_{p,\nu}(z),\widetilde{L}_{q,\nu}(z)   \right\rbrace_{P.B.}&=&i(p-q)\,\widetilde{L}_{p+q,\nu}(z)\nonumber\,.
\ee
It is important to notice that the Virasoro are independent. In order to see this we compute the brackets between them and find that 
\be
\left\lbrace L_{p,\nu}(z),\widetilde{L}_{q,\nu}(z)  \right\rbrace_{P.B.}&=&0\,.
\ee
The explicit calculation is given in the Appendix \ref{Appen:ViraSums}.

It is not strange to find that in fact the centreless Virasoro algebra is satisfied for this theory because the definition given in Eq.~\eqref{VirasoroLtilde} and~\eqref{VirasoroL} is nothing more that a Fourier decomposition (at equal $z$) of the algebra of the Virasoro constraints \eqref{our-EOM-sym-2} in the frational conformal gauge, given by
\be
\left\lbrace T_{--}(z,\sigma),T_{--}(z,\sigma') \right\rbrace_{P.B.}&=&2\pi\Big[T_{--}(z,\sigma)+T_{--}(z,\sigma')\Big]\,\partial_\sigma \,\delta(\sigma-\sigma')\,;\nonumber\\
\left\lbrace T_{++}(z,\sigma),T_{++}(z,\sigma') \right\rbrace_{P.B.}&=&-2\pi\Big[T_{++}(z,\sigma)+T_{++}(z,\sigma')\Big]\,\partial_\sigma \,\delta(\sigma-\sigma')\,;\nonumber\\
\left\lbrace T_{++}(z,\sigma),T_{--}(z,\sigma') \right\rbrace_{P.B.}&=&0\,.
\ee

\section{Residual symmetry and Asymptotic Symmetry}\label{Sec5}

In this section we focus on the study of the in-homogeneous continuity equation in \eqref{conservationlaw}, and its respective associated symmetry. 

In the usual bosonic string theory, after fixing the coformal gauge, the conservation laws  $\nabla^a T_{ab}=0$ lead to residual symmetries when they are written in light-cone coordinates on the WS. In Section \ref{EneMomentTenso} we saw that for the fractional bosonic strings we have  
\be\label{Continuity2}
\nabla^a T_{a1}=0\quad\mbox{and}\quad \nabla^a T_{a0}=g_\alpha(\tau,\sigma)\,,
\ee
with
\be\label{galpha}
g_{\alpha}(\tau,\sigma)\equiv(1-\alpha)\,\partial_0\Big[\ln(t-\tau)\Big]\,T_{00}\,.
\ee

We have shown in the Appendix \ref{Appen:ContiEqu} that the Fractional Bosonic String solution satisfies the equation $\nabla^a T_{a1}=0$. So, is it the second equation also satisfied? The answer is affirmative and to see this we write 
\be\label{gnu}
g_{\nu}(\tau,\sigma)=\frac{(2\nu-1)}{\Gamma(2-2\nu)}\,\mathcal{F}_\nu(z,\sigma),
\ee
with $a=t-\tau$, $\alpha=2-2\nu$, and 
\be
\mathcal{F}_\nu(z,\sigma)&=&z^{2\nu-2}|| \bm{\alpha}_0||^2-\frac{i}{2}\sum_{m\neq0}\texttt{Sgn}(m)\,\mathcal{E}_{\nu-1}^{m}\left(\bm{\alpha}_m\cdot \bm{\alpha}_0\,e^{im\sigma} +\widetilde{\bm{\alpha}}_m\cdot \bm{\alpha}_0\,e^{-im\sigma}\right)-\nonumber\\
&&-\,\frac{1}{4\,z}\sum_{m,n\neq0}\left[ \mathcal{Q}_\nu^{mn}(z)\left(\bm{\alpha}_m\cdot \bm{\alpha}_n\,e^{i(m+n)\sigma}+\widetilde{\bm{\alpha}}_m\cdot\widetilde{\bm{\alpha}}_n\,e^{-i(m+n)\sigma}\right)-\right.\nonumber\\
&&\left.- \,\mathcal{R}_\nu
^{mn}(z)\left(\widetilde{\bm{\alpha}}_m\cdot \bm{\alpha}_n\,e^{-i(m-n)\sigma}+\,\bm{\alpha}_m\cdot\widetilde{\bm{\alpha}}_n\,e^{i(m-n)\sigma}\right)
\right]\,.
\ee
This can be prove by using the derivatives of the embedding map in \eqref{Xdotsquare} and \eqref{Xprimesquare} together with the expressions of $ \mathcal{Q}_\nu^{mn}(z)$ and $ \mathcal{R}_\nu^{mn}(z)$ in the Appendix \ref{Appen:Prop}. Now we need to verify  whether the energy momentum tensor in Eq.\eqref{T++full} and \eqref{T--full}, indeed satisfy the inhomogeneous conservation law in \eqref{Continuity2}.
\begin{proof}
The equation $\nabla^a T_{a0}$ can be written as
\be
\nabla^a T_{a0}= \partial_+\,T_{--}+\partial_-\,T_{++}=\frac{\partial}{\partial z}\left[T_{00}\right]+\frac{\partial}{\partial \sigma}\left[T_{01}\right]\,.
\ee
Using the explicit form of $T_{00}$ and $T_{01}$, we can write
\be
\frac{\partial}{\partial \sigma}\left[T_{01}\right]&=&\frac{1}{4\,\Gamma(2-2\nu)}\left\lbrace 2i\sum_{m\neq0}m\,\mathcal{E}_{\nu}^{m}\left(\bm{\alpha}_m\cdot \bm{\alpha}_0\,e^{im\sigma} +\widetilde{\bm{\alpha}}_m\cdot \bm{\alpha}_0\,e^{-im\sigma}\right)+\right.\nonumber\\
&&+\, \sum_{m,n\neq0} \left[
(m+n)\,\mathcal{X}_\nu^{mn}(z)\,\left(\bm{\alpha}_m\cdot \bm{\alpha}_n\,e^{i(m+n)\sigma}+\widetilde{\bm{\alpha}}_m\cdot\widetilde{\bm{\alpha}}_n\,e^{-i(m+n)\sigma}\right)-\right.\nonumber\\
&&\left.\left.-\,(m-n)\mathcal{Y}_\nu^{mn}(z) \left(\widetilde{\bm{\alpha}}_m\cdot \bm{\alpha}_n\,e^{-i(m-n)\sigma}+\,\bm{\alpha}_m\cdot\widetilde{\bm{\alpha}}_n\,e^{i(m-n)\sigma}\right)
\right]\right\rbrace\nonumber\,;\\
\frac{\partial}{\partial z}\left[T_{00}\right]&=&\frac{1}{4\,\Gamma(2-2\nu)}\left\lbrace 4(2\nu-1)\,z^{2\nu-2\,}||\bm{\alpha}_0||^2-\right.\nonumber\\
&&-\,2i\sum_{m\neq0}\frac{\partial}{\partial z}\left(z\,\texttt{Sgn}(m)\mathcal{E}_{\nu-1}^{m}\right)\left(\bm{\alpha}_m\cdot \bm{\alpha}_0\,e^{im\sigma} +\widetilde{\bm{\alpha}}_m\cdot \bm{\alpha}_0\,e^{-im\sigma}\right)+\nonumber\\
&&+\, \sum_{m,n\neq0} \left[
\frac{\partial}{\partial z}\mathcal{R}_\nu^{mn}(z)\,\left(\bm{\alpha}_m\cdot \bm{\alpha}_n\,e^{i(m+n)\sigma}+\widetilde{\bm{\alpha}}_m\cdot\widetilde{\bm{\alpha}}_n\,e^{-i(m+n)\sigma}\right)-\right.\nonumber\\
&&\left.\left.-\,\frac{\partial}{\partial z}\mathcal{Q}_\nu^{mn}(z) \left(\widetilde{\bm{\alpha}}_m\cdot \bm{\alpha}_n\,e^{-i(m-n)\sigma}+\,\bm{\alpha}_m\cdot\widetilde{\bm{\alpha}}_n\,e^{i(m-n)\sigma}\right)
\right]\right\rbrace\nonumber\,.
\ee
Then, using the relations \eqref{A8}, \eqref{A11}, and \eqref{A12}, we recover the equation \eqref{gnu}.
\end{proof}
Once we have verified our solution we want to know whether there is a way to recover the homogeneous conservation law associated to $\tau$-diffeomorphism invariance. In order see this we need to find the solutions of the equation $g_{\nu}(\tau,\sigma)=0$. 

It is clear that if we take the limit $\nu\rightarrow\frac{1}{2}$ we recover the homogeneous equation but there might be another way to regain this broken symmetry, that is searching when the function $\mathcal{F}_\nu(z,\sigma)$ vanish. In fact if we take the asymptotic approximation $z\gg |\nu^2-\frac{1}{4}|$, the function $\mathcal{F}_\nu(z,\sigma)$ can be written as
\be
\mathcal{F}_\nu(z,\sigma)&\sim&z^{2\nu-2}|| \bm{\alpha}_0||^2+\frac{1}{2}\sum_{m\neq0}z^{\nu-\frac{3}{2}}\left(\bm{\alpha}_m\cdot \bm{\alpha}_0\,e^{im\sigma} +\widetilde{\bm{\alpha}}_m\cdot \bm{\alpha}_0\,e^{-im\sigma}\right)\,e^{-imz}\,e^{-i\,\delta_m(\nu)}+\nonumber\\
&&+\,\frac{1}{2\,z}\sum_{m,n\neq0} \left(\widetilde{\bm{\alpha}}_m\cdot \bm{\alpha}_n\,e^{-i(m-n)\sigma}+\,\bm{\alpha}_m\cdot\widetilde{\bm{\alpha}}_n\,e^{i(m-n)\sigma}\right)e^{-i(m+n)z}\,e^{-i\,[\delta_m(\nu)+\delta_n(\nu)]}\,,\nonumber
\ee
with $\delta_m(\nu)=\frac{\pi}{2}(\nu-\frac{1}{2})\,\texttt{Sgn}(m)$. Therefore, for $\frac{1}{2}<\nu<1$ and $z\gg0$, $\mathcal{F}_\nu(z,\sigma)\rightarrow0$. The limit $z=(t-\tau)\gg0$ can be understood in two different manners: either taking $t\rightarrow+\infty$ or $\tau\rightarrow-\infty$. Both ways take us to the $\tau$-boundary of our theory, where the solution compatible with $t\rightarrow+\infty$ requires also $\nu\rightarrow\frac{1}{2}$.  Therefore, we can clearly see that in the $\tau$-boundary we recover our symmetry.

\section{Conclusion}

After reviewing the concepts of Fractional Bosonic Strings we have introduced the proper definition of the Fractional Virasoro Operators. Then, we have compute the algebra of this objects. Initially, it was not clear whether this new operators would satisfy the centreless Virasoro algebra (also called the Witt Algebra). Once, we checked that the algebra of conformal symmetry was fulfilled, then it was evident that our theory would be invariant under the conformal transformation. 

One of the most interesting facts in this work is the new Fourier decompositions of the Virasaro operators, that we called Fractional Virasoros, which in fact is a new different representation of the Virasoro operators.

Besides finding explicitly that the conformal algebra is satisfied, the presence of asymptotic symmetries was an unexpected result. Asymptotic symmetries has been a branch of gravity and gauge field theory study for quite long time one of the most recent reviews is given~\cite{Strominger:2017zoo}. The early works of asymptotic symmetries were done in gravity. The goal of this idea was to find a subgroup of diffeomorphism of asymptotically flat space-times that act non-trivially on the asymptotic data. In particular, for an asymptotically flat black hole, a boost and a translation are diffeomorphisms that should certainly be allowed but must be non-trivial, because the first one changes the energy and the second one move the black hole to a different place. The group found it in this early work was called BMS group and it is an infinite-dimensional group, that contains the  finite-dimensional Poincar\'e group as a subgroup but has an additional infinity of generators known as "super-translations``. In our case we do not have such group, instead in the $\tau$-boundary of our theory we found out that the conservation law associated to the $\tau$-diffeomorphism is recover. Proving that our theory show $\tau$-diffeormorphism invariance as an asymptotic symmetry.

Finally to conclude, we just remarked that still there is an great amount of work to do in this theory. For example, the inclusion of fermionic degrees of freedom to the theory, to properly represent the baryonic degrees of freedom that we can find in the Universe. As well as properly discuss the inclusion of a cosmological constant and how would it change the equation of motion of the intrinsic metric, and therefore the fractional bosonic solution.

\section*{Acknowledgments}
I am grateful to Lorenz Schlechter, Max Brinkmann, Matthias Traube, and Daniel Kl\"awer, for interesting discussions on conformal field theory. Also I would like to thanks Matteo Capozi and Edoardo Vitagliano for patience discussiosn about field theory. I would like to thanks Andrea Giusti for corrections to the manuscript and guidance in the fractional calculus topic. Lastly, I would like to thanks Prof. Ralph Blumenhagen for agree to be my host and the  Max-Planck-Institut f\"ur Physik for its hospitality and support during the development of this work. This research was partially supported by the Marco Polo scholarship given by the Universit\`a di Bologna and the DAAD Research Grants-Short-Term Grants, 2018 (57381332).

\appendix

\section{Extra terms in $S_{P}$}\label{Appen:Extraterms}

In this appendix we want to see the effects of the terms \eqref{Act1} and \eqref{Act2} on the equations of motions of the time-fractional Polyakov action.

The full action can be written as
	\be \label{fullaction}
	S_{\alpha}\equiv-\frac{1}{4\pi\alpha'\,\Gamma(\alpha)}\int_{0}^{2\pi}d\sigma\,\int_{-\infty}^{t}(t-\tau)^{\alpha-1}\,d\tau\,\sqrt{-h}\left[h^{ab}(\tau,\sigma)\,\partial_a \bm{X}(\tau,\sigma) \cdot \partial_b \bm{X} (\tau,\sigma) +\right.\nonumber\\
\left.	+4\pi\alpha'\,\lambda_1+2\alpha'\lambda_2\,R^{(2)} \right].
	\ee
Therefore, the variation of the action respect to te metric $h$ is given by
		\be \label{variation}
	\delta S_{\alpha}=-\int_{0}^{2\pi}d\sigma\,\int_{-\infty}^{t}\,d\tau\,\sqrt{-h}\left[T_{ab}\,\delta h^{ab}-\frac{\lambda_1\,(t-\tau)^{\alpha-1}}{2\,\Gamma(\alpha)}\,h_{ab}\,\delta h^{ab}+ \right.\nonumber\\
\left.	+\frac{\lambda_2\,(t-\tau)^{\alpha-1}}{2\,\Gamma(\alpha)}\left(R^{(2)}_{ab}-\frac{R^{(2)}}{2}\,h_{ab}+\nabla_c \,V^c\right) \right],
	\ee
Here, we have used several properties such as
\be
\delta h&=& h\,h^{ab}\delta h_{ab}\\
h^{ab}\delta h_{ab}&=&-h_{ab}\delta h^{ab}\\
\nabla_c h_{ab}&=&0\\
\delta R^{(2)}_{ab}&=& \nabla_c\left(\delta \Gamma_{ba}^c\right)-\nabla_b\left(\delta \Gamma_{ca}^c\right)\,,
\ee
where $\nabla_a$ is the covariant derivative in 2-dimensions. The $V^c$ in \eqref{variation} is defined as
\be
V^c=h^{ab} \,\delta\Gamma_{ba}^c-h^{ac}\,\delta \Gamma_{da}^d.
\ee	
We can see in \eqref{variation} that the term with $\lambda_1$ give us an additional cosmological constant term in our equations of motion. It is easy to prove that 2-dimensions $R^{(2)}_{ab}=\frac{R^{(2)}}{2}\,h_{ab}$, telling us that the only $\lambda_2$ contribution is given by
\be
\delta S_{\alpha,2}=-\frac{\lambda_2}{2\,\Gamma(\alpha)}\int_{0}^{2\pi}d\sigma\,\int_{-\infty}^{t}\,d\tau(t-\tau)^{\alpha-1}\sqrt{-h} \, \nabla_c \,V^c.
\ee
In the limit $\alpha\rightarrow1$, this term becomes a total derivative and does not contribute to the equations of motion. However, in this case there is a contribution term of the form
 \be
\delta S_{\alpha,2}=-\frac{\lambda_2}{2\,\Gamma(\alpha)}\int_{0}^{2\pi}d\sigma\,\int_{-\infty}^{t}\,d\tau\sqrt{-h} \, V^c\partial_c \left[(t-\tau)^{\alpha-1}\right].
\ee
So we can only say that our solution of Fractional Bosonic Strings is valid for $\lambda_2=0$, which implies automatically $\lambda_1=0$.
\section{Properties of $\mathcal{E}_{\nu}^m(z)$}\label{Appen:Prop}

In this appendix we prove some useful properties of the $\mathcal{E}_\nu^m(z)$ function using the well known properties of the Hankel functions in [ref].
\begin{enumerate}
\item Hankel functions being a linear combination of the solution of the Bessel function satisfy the Bessel equation
\be
x^2\,\frac{d^2}{dx^2}\mathcal{C}_\alpha (x)+x\,\frac{d}{dx}\mathcal{C}_\alpha (x)+(x^2-\alpha^2)\,\mathcal{C}_\alpha (x)=0,
\ee
with $\mathcal{C}_\alpha (x)=\{ J_\alpha(x),\,Y_\alpha(x),\,H^{(1)}_\alpha(x),\,H^{(2)}_\alpha(x)\}$. It is straightforward to prove the recurrence relation
\be\label{Hankrecurencerela}
\frac{2\alpha}{x}\,\mathcal{C}_\alpha (x)=\mathcal{C}_{\alpha -1}(x)+\mathcal{C}_{\alpha+1} (x).
\ee
Using the above equation and the explicit form of $\mathcal{E}_{\nu}^m(z)$ in terms of Hankel functions, we can write
\be\label{Recurre1}
\mathcal{E}_{\nu}^m(z)=-\frac{m\,\texttt{Sgn}(m)}{2\nu}\,\left[z^2\,\mathcal{E}_{\nu-1}^m(z)+\mathcal{E}_{\nu+1}^m(z)\right].
\ee
Another useful relation is given by shifting $\nu\rightarrow\nu-1$
\be\label{Recurre2}
z^2\,\mathcal{E}_{\nu-2}^m(z)=\frac{(2-2\nu)}{m}\,\texttt{Sgn}(m)\,\mathcal{E}_{\nu-1}^m(z)-\mathcal{E}_{\nu}^m(z).
\ee
\item An important combination of $\mathcal{E}_\nu^m(z)$ is given by
\be\label{Erelation}
\mathcal{E}_\nu^{m}(z) \, \mathcal{E}_{\nu-1}^{-m}(z)-\mathcal{E}_\nu^{-m}(z) \, \mathcal{E}_{\nu-1}^{m}(z) =-2i\,z^{2\nu-2}\,\texttt{Sgn}(m)\,,
\ee
which appears in the computation of $L_{0,\nu}$ ($\widetilde{L}_{0,\nu}$) operators and also in the calculation of the $\alpha(\widetilde{\alpha})$-Algebra.
\begin{proof}
We can write the products as
\be
\mathcal{E}_\nu^{m}(z) \, \mathcal{E}_{\nu-1}^{-m}(z)=\frac{\pi \,|m|}{2}\,z^{2\nu-1}\left\{
\begin{aligned}
&  H ^{(1)}_{-\nu} (|m| z)\,H ^{(2)}_{-\nu+1} (|m| z)  \, , \,\, m \in \mathbb{Z} ^- \, , \\
& H ^{(2)}_{-\nu} (|m| z)\,H ^{(1)}_{-\nu+1} (|m| z) \, , \,\, m \in \mathbb{N} \, , \\
 \end{aligned}
\right.\,;\nonumber\\[2mm]
\mathcal{E}_\nu^{-m}(z) \, \mathcal{E}_{\nu-1}^{m}(z)=\frac{\pi \,|m|}{2}\,z^{2\nu-1}\left\{
\begin{aligned}
&  H ^{(2)}_{-\nu} (|m| z)\,H ^{(1)}_{-\nu+1} (|m| z)  \, , \,\, m \in \mathbb{Z} ^- \, , \\
& H ^{(1)}_{-\nu} (|m| z)\,H ^{(2)}_{-\nu+1} (|m| z) \, , \,\, m \in \mathbb{N} \, , \\
 \end{aligned}
\right.\,.\nonumber
\ee
Therefore, the difference in \eqref{Erelation} can be written as
\be
 \mathcal{E}_\nu^{m}(z) \, \mathcal{E}_{\nu-1}^{-m}(z)-\mathcal{E}_\nu^{-m}(z) \, \mathcal{E}_{\nu-1}^{m}(z) =\frac{\pi \,|m|}{2}\,z^{2\nu-1}\,\texttt{Sgn}(m) \,W[H ^{(1)}_{-\nu} (|m| z),H ^{(2)}_{-\nu} (|m| z)]\,,\nonumber\\
 \label{SumEmE-m}
\ee
where $W[\cdot,\cdot]$ is the Wronskian. Also we now that
\be
W[H ^{(1)}_{-\nu} (x),H ^{(2)}_{-\nu} (x)]=-\frac{4\,i}{\pi x}\,.\nonumber
\ee
Replacing this relation in \eqref{SumEmE-m}, we find
\be\label{Impotrela}
\mathcal{E}_\nu^{m}(z) \, \mathcal{E}_{\nu-1}^{-m}(z)-\mathcal{E}_\nu^{-m}(z) \, \mathcal{E}_{\nu-1}^{m}(z) =-2i\,z^{2\nu-2}\,\texttt{Sgn}(m)\,.
\ee
\end{proof}

\item Also it has been proven in [fracto strings] the following differential equation
\be \label{relaE1}
\frac{\partial}{\partial z}\mathcal{E}_\nu^{m} ( z) = - m\,\texttt{Sgn}(m) \, z \, \mathcal{E}_{\nu-1}^{m} ( z).
\ee
In order to prove \eqref{relaE1},  the recurrence relation of the Hankel function
\be
\frac{1}{x}\frac{\partial}{\partial x}\left[x^{-\alpha}\, \mathcal{C}_\alpha (x)\right]=-x^{-\alpha-1}\,\mathcal{C}_{\alpha+1}(x)\,,
\ee
was used. Another recurrence relation that can be use to decrease the index $\alpha$ to $\alpha-1$, and is given by
\be
\frac{1}{x}\frac{\partial}{\partial x}\left[x^{\alpha}\, \mathcal{C}_\alpha (x)\right]=x^{\alpha-1}\,\mathcal{C}_{\alpha-1}(x)\,.
\ee
Therefore, we can write the following differential equation
\be\label{relaE2}
\frac{\partial}{\partial z}\left[z^{-2\nu}\,\mathcal{E}_\nu^{m} ( z) \right]= m\,\texttt{Sgn}(m) \, z^{-1-2\nu} \, \mathcal{E}_{\nu+1}^{m} ( z)\,.
\ee
\begin{proof}
Using the explicit form of $\mathcal{E}_\nu^m(z)$
\be\mathcal{E}_\nu^{m}(  z) =
\sqrt{\frac{\pi\,|m|}{2}} \, z^\nu \times
\left\{
\begin{aligned}
& H ^{(1)}_{-\nu} (|m| z) \, , \,\, m \in \mathbb{Z} ^- \, , \\
& H ^{(2)}_{-\nu} (|m| z) \, , \,\, m \in \mathbb{N} \, , \\
 \end{aligned}
\right.\,,
\ee
we can write
\be
\frac{\partial}{\partial z}\left[z^{-2\nu}\,\mathcal{E}_\nu^{m} ( z) \right]&=&\sqrt{\frac{\pi\,|m|}{2}} 
\times\left\{
\begin{aligned}
& \frac{\partial}{\partial z} \left[z^{-\nu}  H ^{(1)}_{-\nu} (|m| z)\right] \, , \,\, m \in \mathbb{Z} ^- \, , \\
&  \frac{\partial}{\partial z} \left[ z^{-\nu} H ^{(2)}_{-\nu} (|m| z)\right] \, , \,\, m \in \mathbb{N} \, , \\
 \end{aligned}
\right.\,\nonumber\\
&=&\sqrt{\frac{\pi\,|m|}{2}} 
\times\,|m|\,z^{-\nu}\left\{
\begin{aligned}
&  H ^{(1)}_{-\nu-1} (|m| z)\, , \,\, m \in \mathbb{Z} ^- \, , \\
&   H ^{(2)}_{-\nu-1} (|m| z)\, , \,\, m \in \mathbb{N} \, , \\
 \end{aligned}
\right.\,\nonumber\\
&=&|m|\,z^{-1-2\nu}\,\mathcal{E}_{\nu+1}^m(z)\nonumber\\
&=&m\,\texttt{Sgn}(m)\,\,z^{-1-2\nu}\,\mathcal{E}_{\nu+1}^m(z)\,.\nonumber
\ee
\end{proof}
\item  In Section \ref{Sol.Eq.Motion}, we have defined the combinations
\be
E^\nu_m(z)&=&\frac{z^{1/2-\nu}}{2}\left[ \mathcal{E}_{\nu}^{m} (z)+iz\,\texttt{Sgn}(m)\,\mathcal{E}_{\nu-1}^{m} (z) \right]\, ,\\
\widetilde{E}^\nu_m(z)&=&\frac{z^{1/2-\nu}}{2}\left[ \mathcal{E}_{\nu}^{m} (z)-iz\,\texttt{Sgn}(m)\,\mathcal{E}_{\nu-1}^{m} (z) \right]\,.
\ee
One can write the products as 
\be \label{EEtilde}
E^\nu_m(z)\,\widetilde{E}^\nu_n(z)=\frac{\mathcal{Q}_\nu^{mn}(z)+i\,\mathcal{Y}_\nu^{mn}(z)}{4}&,& E^\nu_n(z)\,\widetilde{E}^\nu_m(z)=\frac{\mathcal{Q}_\nu^{mn}(z)-i\,\mathcal{Y}_\nu^{mn}(z)}{4}\,,\\
&\mbox{and }&\nonumber\\\label{EE}
E^\nu_m(z)\,E^\nu_n(z)=\frac{\mathcal{R}_\nu^{mn}(z)+i\,\mathcal{X}_\nu^{mn}(z)}{4}&,& \widetilde{E}^\nu_m (z)\widetilde{E}^\nu_n(z)=\frac{\mathcal{R}_\nu^{mn}(z)-i\,\mathcal{X}_\nu^{mn}(z)}{4}\,.
\ee
Here, we have introduced
\be\label{X}
\mathcal{X}_\nu^{mn}(z)&\equiv& z^{2-2\nu}\left(\texttt{Sgn} (m) \,\mathcal{E}_{\nu-1}^{m}  \,\mathcal{E}_{\nu}^{n}+\texttt{Sgn} (n) \,\mathcal{E}_{\nu-1}^{n}  \,\mathcal{E}_{\nu}^{m}\right)\,;\\[2mm]
\mathcal{Y}_\nu^{mn}(z)&\equiv& z^{2-2\nu}\left(\texttt{Sgn} (m) \,\mathcal{E}_{\nu-1}^{m}  \,\mathcal{E}_{\nu}^{n}-\texttt{Sgn} (n) \,\mathcal{E}_{\nu-1}^{n}  \,\mathcal{E}_{\nu}^{m}\right)\,;\\[2mm]
\mathcal{Q}_\nu^{mn}(z)&\equiv&z^{1-2\nu}\left(\mathcal{E}_{\nu}^{m}\mathcal{E}_{\nu}^{n}+z^2\,\texttt{Sgn} (m)\,\texttt{Sgn} (n) \,\mathcal{E}_{\nu-1}^{m}  \,\mathcal{E}_{\nu-1}^{n} \right)\,;\\[2mm]
\mathcal{R}_\nu^{mn}(z)&\equiv&z^{1-2\nu}\left(\mathcal{E}_{\nu}^{m}\mathcal{E}_{\nu}^{n}-z^2\,\texttt{Sgn} (m)\,\texttt{Sgn} (n) \,\mathcal{E}_{\nu-1}^{m}  \,\mathcal{E}_{\nu-1}^{n} \right)\,.
\ee
Using \eqref{Recurre1}, \eqref{Recurre2}, \eqref{relaE1}, and \eqref{relaE2}, is not difficult to prove the following relations
\be \label{A8}
\frac{\partial}{\partial z}\left(z\,\texttt{Sgn}(m)\,\mathcal{E}_{\nu-1}^m\right)&=&(2\nu-1)\,\texttt{Sgn}(m)\,\mathcal{E}_{\nu-1}^m+m\,\mathcal{E}_{\nu}^m\,;\\[2mm]\label{A9}
\frac{\partial}{\partial z}\mathcal{X}_\nu^{mn}(z)&=&(m+n)\,\mathcal{R}_\nu^{mn}(z)\,;\\[2mm]\label{A10}
\frac{\partial}{\partial z} \mathcal{Y}_\nu^{mn}(z)&=&(m-n)\,\mathcal{Q}_\nu^{mn}(z)\,;\\[2mm]\label{A11}
\frac{\partial}{\partial z}\mathcal{R}_\nu^{mn}(z)&=&\frac{(1-2\nu)}{z}\,\mathcal{Q}_\nu^{mn}(z)-(m+n)\,\mathcal{X}_\nu^{mn}(z)\,;\\[2mm]\label{A12}
\frac{\partial}{\partial z}\mathcal{Q}_\nu^{mn}(z)&=&\frac{(1-2\nu)}{z}\,\mathcal{R}_\nu^{mn}(z)-(m-n)\,\mathcal{Y}_\nu^{mn}(z)\,.
\ee

\end{enumerate}

\section{Continuity equation}\label{Appen:ContiEqu}

The purpose of this appendix is to show the solution \eqref{eq-sol}  satisfy the continuity equation \eqref{continuiyEqFracBos}. It was mentioned in Section \ref{EneMomentTenso} that the energy momentum tensor satisfy the continuity equation
\be
\partial_+\,T_{--}-\partial_-\,T_{++}=0\,,
\ee
in the fracitonal conformal gauge with $f(\tau,\sigma)=1$. This equation can also be written as
\be
\frac{\partial}{\partial z}\left[T_{01}\right]+\frac{\partial}{\partial \sigma}\left[T_{00}\right]=0\,,
\ee
where $z=t-\tau$. Replacing the explicit expressions for $T_{01}$ and $T_{00}$, we can write
\be\label{ContinuityAppendix}
\frac{\partial}{\partial z}\left[z^{1-2\nu}\,\Big(2\,\dot{\bm{X}}\cdot\bm{X}'\Big)\right]+\frac{\partial}{\partial \sigma}\left[z^{1-2\nu}\Big(||\dot{\bm{X}}||^2+||\bm{X}'||^2\Big)\right]=0\,,
\ee

The square of  the derivatives of the solution \eqref{Xdot} and \eqref{Xprime} are given by
\be 
&&||\dot{\bm{X}}||^2= \frac{\alpha'}{2}\left[ 4\,z^{4\nu-2}\,|| \bm{\alpha}_0||^2-z^2\sum_{m,n\neq0} \left(\bm{\alpha}_m\cdot \bm{\alpha}_n\,e^{i(m+n)\sigma}+\widetilde{\bm{\alpha}}_m\cdot \bm{\alpha}_n\,e^{-i(m-n)\sigma}\right.\right.\nonumber\\
&&\left.\left.+ \,\bm{\alpha}_m\cdot\widetilde{\bm{\alpha}}_n\,e^{i(m-n)\sigma}+\widetilde{\bm{\alpha}}_m\cdot\widetilde{\bm{\alpha}}_n\,e^{-i(m+n)\sigma}\right)\texttt{Sgn} (m)\texttt{Sgn} (n) \,\mathcal{E}_{\nu-1}^{m} (z) \,\mathcal{E}_{\nu-1}^{n}(z)-\right.\nonumber\\[2mm]\label{Xdotsquare}
&&\left.-\,2iz^{2\nu}\sum_{m\neq0}\left(\bm{\alpha}_m\cdot \bm{\alpha}_0\,e^{im\sigma} +\widetilde{\bm{\alpha}}_m\cdot \bm{\alpha}_0\,e^{-im\sigma}\right)\texttt{Sgn} (m) \,\mathcal{E}_{\nu-1}^{m} (z)\right];\\[2mm]
&&||\bm{X}'||^2 = \frac{\alpha'}{2}\left[\sum_{m,n\neq0}\left(\bm{\alpha}_m\cdot \bm{\alpha}_n\,e^{i(m+n)\sigma}-\widetilde{\bm{\alpha}}_m\cdot \bm{\alpha}_n\,e^{-i(m-n)\sigma}-\right.\right.\nonumber\\\label{Xprimesquare}
&&\left.\left.- \,\bm{\alpha}_m\cdot\widetilde{\bm{\alpha}}_n\,e^{i(m-n)\sigma}+\widetilde{\bm{\alpha}}_m\cdot\widetilde{\bm{\alpha}}_n\,e^{-i(m+n)\sigma}\right)\mathcal{E}_{\nu}^{m}(z)  \,\mathcal{E}_{\nu}^{n}(z)\right].
\ee
Also the scalar product is given by
\be\label{XdotXprime}
&&2\,\dot{\bm{X}}\cdot\bm{X}'=\frac{\alpha'}{2} \left[ 2\,z^{2\nu-1}\sum_{m\neq0}\left(\bm{\alpha}_m\cdot \bm{\alpha}_0\,e^{im\sigma} -\widetilde{\bm{\alpha}}_m\cdot \bm{\alpha}_0\,e^{-im\sigma}\right)\mathcal{E}_{\nu}^{m} (z)-\right.\nonumber\\
&&-\,iz\sum_{m,n\neq0} \left(\left[\bm{\alpha}_m\cdot \bm{\alpha}_n\,e^{i(m+n)\sigma}-\widetilde{\bm{\alpha}}_m\cdot\widetilde{\bm{\alpha}}_n\,e^{-i(m+n)\sigma}\right]\left[\texttt{Sgn} (m) \,\mathcal{E}_{\nu-1}^{m} (z) \,\mathcal{E}_{\nu}^{n}(z)+\right.\right.\nonumber\\
&&\left.\left.+\,\texttt{Sgn} (n) \,\mathcal{E}_{\nu-1}^{n} (z) \,\mathcal{E}_{\nu}^{m}(z)\right]+\left[\widetilde{\bm{\alpha}}_m\cdot \bm{\alpha}_n\,e^{-i(m-n)\sigma}- \,\bm{\alpha}_m\cdot\widetilde{\bm{\alpha}}_n\,e^{i(m-n)\sigma}\right]\left[\texttt{Sgn} (m) \,\mathcal{E}_{\nu-1}^{m} (z) \,\mathcal{E}_{\nu}^{n}(z)\right.\right.\nonumber\\
&&\left.\left.-\texttt{Sgn} (n) \,\mathcal{E}_{\nu-1}^{n} (z) \,\mathcal{E}_{\nu}^{m}(z)\right]
\right).
\ee
Therefore, we can write
\be
\frac{\partial}{\partial z}\left[z^{1-2\nu}\,\Big(2\,\dot{\bm{X}}\cdot\bm{X}'\Big)\right]&=&\frac{\alpha'}{2}\left[2\,\sum_{m\neq0}\left(\bm{\alpha}_m\cdot \bm{\alpha}_0\,e^{im\sigma} -\widetilde{\bm{\alpha}}_m\cdot \bm{\alpha}_0\,e^{-im\sigma}\right)\frac{\partial\mathcal{E}_{\nu}^{m}}{\partial z}-\right.\nonumber\\
&&\left.-\,i\sum_{m,n\neq0} \left\lbrace
\frac{\partial}{\partial z}\mathcal{X}_\nu^{mn}(z)\,
\left(\bm{\alpha}_m\cdot \bm{\alpha}_n\,e^{i(m+n)\sigma}-\widetilde{\bm{\alpha}}_m\cdot\widetilde{\bm{\alpha}}_n\,e^{-i(m+n)\sigma}\right)+\right.\right.\nonumber\\
&&\left.\left.+\,\frac{\partial}{\partial z}\mathcal{Y}_\nu^{mn}(z)
\left(\widetilde{\bm{\alpha}}_m\cdot \bm{\alpha}_n\,e^{-i(m-n)\sigma}- \,\bm{\alpha}_m\cdot\widetilde{\bm{\alpha}}_n\,e^{i(m-n)\sigma}\right)
\right\rbrace\right]\,;\\
\frac{\partial}{\partial \sigma}\left[z^{1-2\nu}\Big(||\dot{\bm{X}}||^2+||\bm{X}'||^2\Big)\right]&=&\frac{\alpha'}{2}\left[2z\sum_{m\neq0}\left(\bm{\alpha}_m\cdot \bm{\alpha}_0\,e^{im\sigma} -\widetilde{\bm{\alpha}}_m\cdot \bm{\alpha}_0\,e^{-im\sigma}\right)\,m\,\texttt{Sgn}(m)\mathcal{E}_{\nu-1}^{m}+\right.\nonumber\\
&&+\,i\sum_{m,n\neq0} \left\lbrace
(m+n)\,\mathcal{R}_\nu^{mn}(z)\left(\bm{\alpha}_m\cdot \bm{\alpha}_n\,e^{i(m+n)\sigma}-\widetilde{\bm{\alpha}}_m\cdot\widetilde{\bm{\alpha}}_n\,e^{-i(m+n)\sigma}\right)+\right.\nonumber\\
&&\left.\left.+\,(m-n)\,\mathcal{Q}_\nu^{mn}(z)\left(\widetilde{\bm{\alpha}}_m\cdot \bm{\alpha}_n\,e^{-i(m-n)\sigma}- \,\bm{\alpha}_m\cdot\widetilde{\bm{\alpha}}_n\,e^{i(m-n)\sigma}\right)
\right\rbrace\right]\,.
\ee
Therefore, we can clearly see that the continuity equation \eqref{ContinuityAppendix} implies the relations \eqref{relaE1}, \eqref{A9}, and \eqref{A10}. Finding an extra cross-check of the solution \eqref{eq-sol}.

\section{Virasoro sums}\label{Appen:ViraSums}
In this Appendix we focus on computing the Possion Brackets of he Virasoro Operators, given by
\be
\left\lbrace L_{p,\nu}(z),L_{q,\nu}(z)  \right\rbrace_{P.B.}&=&\frac{1}{2\,\Gamma(2-2\nu)}\sum_{n\in\mathbb{Z}}\Big(\widetilde{E}_{q-n}^\nu\,\widetilde{E}_n^\nu\,\left\lbrace L_{p,\nu}(z),\bm{\alpha}_{q-n}\cdot\bm{\alpha}_n\right\rbrace_{P.B.}-\nonumber\\
&&-\,E_{n-q}^\nu\,\widetilde{E}_{n}^\nu\,\left\lbrace L_{p,\nu}(z),\widetilde{\bm{\alpha}}_{n-q}\cdot\bm{\alpha}_n\right\rbrace_{P.B.}-E_{n}^\nu\,\widetilde{E}_{n+q}^\nu\,\left\lbrace L_{p,\nu}(z),\bm{\alpha}_{n+q}\cdot\widetilde{\bm{\alpha}}_n\right\rbrace_{P.B.}\nonumber\\
&&+\,E_{n}^\nu\,E_{-n-q}^\nu\,\left\lbrace L_{p,\nu}(z),\widetilde{\bm{\alpha}}_{-n-q}\cdot\widetilde{\bm{\alpha}}_n\right\rbrace_{P.B.}\Big)\,.
\ee
Let us compute term by term in this sum. The first contribution is given by
\be
&\displaystyle{\sum_{n\in\mathbb{Z}}\widetilde{E}_{q-n}^\nu\,\widetilde{E}_n^\nu\,\left\lbrace L_{p,\nu}(z),\bm{\alpha}_{q-n}\cdot\bm{\alpha}_n\right\rbrace_{P.B.}=-i\sum_{n\in\mathbb{Z}}\widetilde{E}_{q-n}^\nu\,\widetilde{E}_n^\nu\,\left\lbrace(q-n)\Big[ \underbrace{ \widetilde{E}_{p+q-n}^\nu\,\widetilde{E}_{n-q}^\nu\,\bm{\alpha} _{p+q-n}\cdot\bm{\alpha} _{n}}_{n=k}-\right.}&\nonumber\\
&\displaystyle{\left.- \underbrace{E_{n-p-q}^\nu\,\widetilde{E}_{n-q}^\nu\,\widetilde{\bm{\alpha}} _{n-p-q}\cdot\bm{\alpha} _{n}}_{n=k}\Big]+n\Big[\underbrace{ \widetilde{E}_{p+n}^\nu\,\widetilde{E}_{-n}^\nu\,\bm{\alpha}_{p+n}\cdot\bm{\alpha}_{q-n}}_{k=n+p}- \underbrace{E_{-n-p}^\nu\,\widetilde{E}_{-n}^\nu\,\widetilde{\bm{\alpha}} _{-p-n}\cdot\bm{\alpha}_{q-n}}_{-k=n+p}\Big]\right\rbrace\,;}&\nonumber\\
&\displaystyle{=\sum_{k\in\mathbb{Z}}\left\lbrace\Big[(k-p)\,\widetilde{E}_{k-p}^\nu\,\widetilde{E}_{p-k}^\nu-(k-q)\,\widetilde{E}_{k-q}^\nu\,\widetilde{E}_{q-k}^\nu\Big]\widetilde{E}_{p+q-k}^\nu\,\widetilde{E}_k^\nu\,\bm{\alpha}_{p+q-k}\cdot\bm{\alpha}_k+\right.}&\nonumber\\
&\displaystyle{\left.+(k-q)\,\widetilde{E}_{k-q}^\nu\,\widetilde{E}_{q-k}^\nu\,\widetilde{E}_k^\nu\, E_{k-p-q}^\nu\,\widetilde{\bm{\alpha}}_{k-p-q}\cdot\bm{\alpha}_k +(k+p)\,\widetilde{E}_{-k-p}^\nu\,\widetilde{E}_{k+p}^\nu\,E_{k}^\nu\,\widetilde{E}_{k+p+q}^\nu\,\widetilde{\bm{\alpha}} _{k}\cdot\bm{\alpha}_{k+p+q}\right\rbrace\,.}&\nonumber
\ee
The next two terms can be written as
\be
&\displaystyle{
\sum_{n\in\mathbb{Z}}E_{n-q}^\nu\,\widetilde{E}_{n}^\nu\,\left\lbrace L_{p,\nu}(z),\widetilde{\bm{\alpha}}_{n-q}\cdot\bm{\alpha}_n\right\rbrace_{P.B.}=-i\sum_{n\in\mathbb{Z}}E_{n-q}^\nu\,\widetilde{E}_n^\nu\,\left\lbrace(n-q)\Big[ \underbrace{ E_{n-p-q}^\nu\,\widetilde{E}_{q-n}^\nu\,\bm{\widetilde{\alpha}} _{n-p-q}\cdot\bm{\alpha} _{n}}_{n=k}-\right.
}&\nonumber\\
&\displaystyle{\left.- \underbrace{E_{q-n}^\nu\,\widetilde{E}_{p+q-n}^\nu\,\bm{\alpha} _{p+q-n}\cdot\bm{\alpha} _{n}}_{n=k}\Big]+n\Big[\underbrace{ \widetilde{E}_{p+n}^\nu\,\widetilde{E}_{-n}^\nu\,\bm{\alpha}_{p+n}\cdot\bm{\widetilde{\alpha}}_{n-q}}_{k=n+p}- \underbrace{E_{-n-p}^\nu\,\widetilde{E}_{-n}^\nu\,\widetilde{\bm{\alpha}} _{-p-n}\cdot\widetilde{\bm{\alpha}}_{n-q}}_{-k=n+p}\Big]\right\rbrace\,;}&\nonumber\\
&\displaystyle{=\sum_{k\in\mathbb{Z}}\left\lbrace\Big[(k-p)\,\widetilde{E}_{k-p}^\nu\,\widetilde{E}_{p-k}^\nu+(k-q)\,E_{k-q}^\nu\,E_{q-k}^\nu\Big]E_{k-p-q}^\nu\,\widetilde{E}_k^\nu\,\widetilde{\bm{\alpha}}_{k-p-q}\cdot\bm{\alpha}_k-\right.}&\nonumber\\
&\displaystyle{\left.-(k-q)\,E_{k-q}^\nu\,E_{q-k}^\nu\,\widetilde{E}_k^\nu\, \widetilde{E}_{p+q-k}^\nu\,\bm{\alpha}_{p+q-k}\cdot\bm{\alpha}_k+(k+p)\,\widetilde{E}_{-k-p}^\nu\,\widetilde{E}_{k+p}^\nu\,E_{k}^\nu\,E_{-k-p-q}^\nu\,\widetilde{\bm{\alpha}} _{k}\cdot\widetilde{\bm{\alpha}}_{-k-p-q}\right\rbrace\,;}&\nonumber
\ee
\be
&\displaystyle{\sum_{n\in\mathbb{Z}}E_{n}^\nu\,\widetilde{E}_{n+q}^\nu\,\left\lbrace L_{p,\nu}(z),\bm{\alpha}_{n+q}\cdot\widetilde{\bm{\alpha}}_n\right\rbrace_{P.B.}=-i\sum_{n\in\mathbb{Z}}E_{n}^\nu\,\widetilde{E}_{n+q}^\nu\,\left\lbrace(n+q)\Big[ \underbrace{ \widetilde{E}_{n+p+q}^\nu\,\widetilde{E}_{-q-n}^\nu\,\bm{\alpha}_{p+q+n}\cdot\widetilde{\bm{\alpha}} _{n}}_{n=k}-\right.}&\nonumber\\
&\displaystyle{\left.- \underbrace{\widetilde{E}_{-q-n}^\nu\,E_{-p-q-n}^\nu\,\widetilde{\bm{\alpha}} _{-p-q-n}\cdot\widetilde{\bm{\alpha}} _{n}}_{n=k}\Big]+n\Big[\underbrace{ E_{n-p}^\nu\,E_{-n}^\nu\,\widetilde{\bm{\alpha}}_{n-p}\cdot\bm{\alpha}_{n+q}}_{k=n-p}-\underbrace{\widetilde{E}_{p-n}^\nu\,E_{-n}^\nu\,\bm{\alpha} _{p-n}\cdot\bm{\alpha}_{n+q}}_{k=p-n}\Big]\right\rbrace\,;}&\nonumber\\
&\displaystyle{=\sum_{k\in\mathbb{Z}}\left\lbrace\Big[(k+p)\,E_{-k-p}^\nu\,E_{p+k}^\nu+(k+q)\,\widetilde{E}_{k+q}^\nu\,\widetilde{E}_{-k-q}^\nu\Big]\widetilde{E}_{k+p+q}^\nu\,E_k^\nu\,\bm{\alpha}_{k+p+q}\cdot\widetilde{\bm{\alpha}}_k-\right.}&\nonumber\\
&\displaystyle{\left.-(k+q)\,\widetilde{E}_{k+q}^\nu\,\widetilde{E}_{-k-q}^\nu\,E_k^\nu\, E_{-p-q-k}^\nu\,\widetilde{\bm{\alpha}}_{-p-q-k}\cdot\widetilde{\bm{\alpha}}_k+(k-p)\,E_{k-p}^\nu\,E_{p-k}^\nu\,\widetilde{E}_{k}^\nu\,\widetilde{E}_{p+q-k}^\nu\,\bm{\alpha} _{p+q-k}\cdot\bm{\alpha}_{k}\right\rbrace\,.}&\nonumber
\ee
Finally the last contribution can be written as
\be
&\displaystyle{\sum_{n\in\mathbb{Z}}E_{n}^\nu\,E_{-n-q}^\nu\,\left\lbrace L_{p,\nu}(z),\widetilde{\bm{\alpha}}_{-n-q}\cdot\widetilde{\bm{\alpha}}_n\right\rbrace_{P.B.}=-i\sum_{n\in\mathbb{Z}}E_{n}^\nu\,E_{-n-q}^\nu\,\left\lbrace-(n+q)\Big[ \underbrace{ E_{-n-p-q}^\nu\,E_{n+q}^\nu\,\widetilde{\bm{\alpha}}_{-n-p-q}\cdot\widetilde{\bm{\alpha}} _{n}}_{n=k}-\right.
}&\nonumber\\
&\displaystyle{\left.- \underbrace{E_{n+q}^\nu\,\widetilde{E}_{n+p+q}^\nu\,\bm{\alpha} _{n+p+q}\cdot\widetilde{\bm{\alpha}} _{n}}_{n=k}\Big]+n\Big[\underbrace{ E_{n-p}^\nu\,E_{-n}^\nu\,\widetilde{\bm{\alpha}}_{n-p}\cdot\widetilde{\bm{\alpha}}_{-n-q}}_{k=n-p}- \underbrace{\widetilde{E}_{p-n}^\nu\,E_{-n}^\nu\,\bm{\alpha} _{p-n}\cdot\widetilde{\bm{\alpha}}_{-n-q}}_{k=p-n}\Big]\right\rbrace\,;}&\nonumber\\
&\displaystyle{=\sum_{k\in\mathbb{Z}}\left\lbrace\Big[(k+p)\,E_{-k-p}^\nu\,E_{p+k}^\nu-\,(k+q)\,E_{k+q}^\nu\,E_{-k-q}^\nu\Big]E_{-k-p-q}^\nu\,E_k^\nu\,\widetilde{\bm{\alpha}}_{-k-p-q}\cdot\widetilde{\bm{\alpha}}_k+\right.
}&\nonumber\\
&\displaystyle{\left.+(k+q)\,E_{k+q}^\nu\,E_{-k-q}^\nu\,E_k^\nu\,\widetilde{E}_{k+p+q}^\nu\,\bm{\alpha}_{k+p+q}\cdot\widetilde{\bm{\alpha}}_k+\,(k-p)\,E_{k-p}^\nu\,E_{p-k}^\nu\,\widetilde{E}_{k}^\nu\,E_{k-p-q}^\nu\,\widetilde{\bm{\alpha}} _{k-p-q}\cdot\bm{\alpha}_{k}\right\rbrace\,.}&\nonumber
\ee
The addition of all these terms leads to
\be
&\displaystyle{
\left\lbrace L_{p,\nu}(z),L_{q,\nu}(z)  \right\rbrace_{P.B.}=\frac{-i}{2\,\Gamma(2-2\nu)}\sum_{k\in\mathbb{Z}}\left\lbrace K_{k-p\,,\,k-q}^\nu(z)\,\widetilde{E}_{p+q-k}^\nu\,\widetilde{E}_{k}^\nu\,\bm{\alpha} _{p+q-k}\cdot\bm{\alpha}_{k}-\right.}&\nonumber\\
&\displaystyle{-\,K^\nu_{k-p\,,\,k-q}(z)\,E_{k-p-q}^\nu\,\widetilde{E}_{k}^\nu\,\widetilde{\bm{\alpha}}_{k-p-q}\cdot\bm{\alpha}_{k}-\,K^\nu_{-k-p\,,\,-k-q}(z)\,\widetilde{E}_{k+p+q}^\nu\,E_{k}^\nu\,\bm{\alpha}_{k+p+q}\cdot\widetilde{\bm{\alpha}}_{k}+}&\nonumber\\
&\displaystyle{\left.+K^\nu_{-k-p\,,\,-k-q}(z)\,E_{-k-p-q}^\nu\,E_{k}^\nu\,\widetilde{\bm{\alpha}}_{-k-p-q}\cdot\widetilde{\bm{\alpha}}_{k}\right\rbrace\,,}&
\ee
where we have defined
\be
K_{m,n}^\nu(z)\equiv\Big[m\,\left(\widetilde{E}_{-m}^\nu\,\widetilde{E}_{m}^\nu-E_{-m}^\nu\,E_{m}^\nu\right)-n\,\left(\widetilde{E}_{-n}^\nu\,\widetilde{E}_{n}^\nu-E_{-n}^\nu\,E_{n}^\nu\right)\Big]\,\quad\forall m,n\in\mathbb{Z}.
\ee
 
For completion, we also compute the Poisson brackets between $L$ and $\widetilde{L}$, which is given by
 \be
\left\lbrace L_{p,\nu}(z),\widetilde{L}_{q,\nu}(z)  \right\rbrace_{P.B.}&=&\frac{1}{2\,\Gamma(2-2\nu)}\sum_{n\in\mathbb{Z}}\Big(E_{-q-n}^\nu\,E_n^\nu\,\left\lbrace L_{p,\nu}(z),\bm{\alpha}_{-q-n}\cdot\bm{\alpha}_n\right\rbrace_{P.B.}-\nonumber\\
&&-\,E_{n-q}^\nu\,\widetilde{E}_{n}^\nu\,\left\lbrace L_{p,\nu}(z),\bm{\alpha}_{n-q}\cdot\widetilde{\bm{\alpha}}_n\right\rbrace_{P.B.}-E_{n}^\nu\,\widetilde{E}_{n+q}^\nu\,\left\lbrace L_{p,\nu}(z),\widetilde{\bm{\alpha}}_{n+q}\cdot\bm{\alpha}_n\right\rbrace_{P.B.}\nonumber\\
&&+\,\widetilde{E}_{n}^\nu\,\widetilde{E}_{q-n}^\nu\,\left\lbrace L_{p,\nu}(z),\widetilde{\bm{\alpha}}_{q-n}\cdot\widetilde{\bm{\alpha}}_n\right\rbrace_{P.B.}\Big)\,.
\ee
We have that the first sum is given by
\begin{eqnarray}
&\displaystyle{\sum_{n\in\mathbb{Z}}E_{-q-n}^\nu\,E_n^\nu\,\left\lbrace L_{p,\nu}(z),\bm{\alpha}_{-q-n}\cdot\bm{\alpha}_n\right\rbrace_{P.B.}=-i\sum_{n\in\mathbb{Z}}E_{-q-n}^\nu\,E_n^\nu\,\left\lbrace(n+q)\Big[\underbrace{E_{n-p+q}^\nu\,\widetilde{E}_{n+q}^\nu\,\widetilde{\bm{\alpha}} _{n+q-p}\cdot\bm{\alpha} _{n}}_{n=k}-\right.}&\nonumber\\
&\displaystyle{\left. -\underbrace{ \widetilde{E}_{p-q-n}^\nu\,\widetilde{E}_{n+q}^\nu\,\bm{\alpha} _{p-q-n}\cdot\bm{\alpha} _{n}}_{n=k}\Big]+n\Big[\underbrace{ \widetilde{E}_{p+n}^\nu\,\widetilde{E}_{-n}^\nu\,\bm{\alpha}_{p+n}\cdot\bm{\alpha}_{-q-n}}_{k=n+p}- \underbrace{E_{-n-p}^\nu\,\widetilde{E}_{-n}^\nu\,\widetilde{\bm{\alpha}} _{-p-n}\cdot\bm{\alpha}_{-q-n}}_{-k=n+p}\Big]\right\rbrace\,;}&\nonumber\\
&\displaystyle{=\sum_{k\in\mathbb{Z}}\left\lbrace\Big[(k-p)\,\widetilde{E}_{p-k}^\nu\,E_{k-p}^\nu\,E_k^\nu\,\widetilde{E}_{p-q-k}-(k+q)\,\widetilde{E}_{k+q}^\nu\,E_{-k-q}^\nu\,\widetilde{E}_{p-q-k}^\nu\,E_k^\nu\Big]\,\bm{\alpha}_{p-q-k}\cdot\bm{\alpha}_k+\right.}&\nonumber\\
&\displaystyle{\left.+(k+q)\,\widetilde{E}_{k+q}^\nu\,E_{-q-k}^\nu\,E_k^\nu\, E_{k-p+q}^\nu\,\widetilde{\bm{\alpha}}_{k-p+q}\cdot\bm{\alpha}_k +(k+p)\,\widetilde{E}_{k+p}^\nu\,E_{-k-p}^\nu\,E_{k}^\nu\,E_{k+p-q}^\nu\,\widetilde{\bm{\alpha}} _{k}\cdot\bm{\alpha}_{k+p-q}\right\rbrace\,.}&\nonumber
\end{eqnarray}
 The second and third contribution are
 \be
&\displaystyle{
\sum_{n\in\mathbb{Z}}E_{n-q}^\nu\,\widetilde{E}_{n}^\nu\,\left\lbrace L_{p,\nu}(z),\bm{\alpha}_{n-q}\cdot\widetilde{\bm{\alpha}}_n\right\rbrace_{P.B.}=-i\sum_{n\in\mathbb{Z}}E_{n-q}^\nu\,\widetilde{E}_n^\nu\,\left\lbrace(n-q)\Big[ \underbrace{ \widetilde{E}_{n+p-q}^\nu\,\widetilde{E}_{q-n}^\nu\,\bm{\alpha} _{n+p-q}\cdot\widetilde{\bm{\alpha} }_{n}}_{n=k}-\right.
}&\nonumber\\
&\displaystyle{\left.- \underbrace{\widetilde{E}_{q-n}^\nu\,E_{-p+q-n}^\nu\,\widetilde{\bm{\alpha}} _{-p+q-n}\cdot\widetilde{\bm{\alpha}} _{n}}_{n=k}\Big]+n\Big[\underbrace{ E_{n-p}^\nu\,E_{-n}^\nu\,\widetilde{\bm{\alpha}}_{n-p}\cdot\bm{\alpha}_{n-q}}_{k=n-p}- \underbrace{E_{-n}^\nu\,\widetilde{E}_{p-n}^\nu\,\bm{\alpha} _{p-n}\cdot\bm{\alpha}_{n-q}}_{k=p-n}\Big]\right\rbrace\,;}&\nonumber\\
&\displaystyle{=\sum_{k\in\mathbb{Z}}\left\lbrace(k-p)\,E_{k-p}^\nu\,\widetilde{E}_{p-k}^\nu\,E_{k+p-q}^\nu\,\widetilde{E}_k^\nu\,\bm{\alpha}_{k+p-q}\cdot\bm{\alpha}_k+(k-q)\,E_{k-q}^\nu\,\widetilde{E}_{q-k}^\nu\,\widetilde{E}_{k+p-q}^\nu\,\widetilde{E}_k^\nu\,\bm{\alpha}_{k+p-q}\cdot\widetilde{\bm{\alpha}}_k-\right.}&\nonumber\\
&\displaystyle{\left.-(k-q)\,E_{k-q}^\nu\,\widetilde{E}_{q-k}^\nu\,\widetilde{E}_k^\nu\, E_{q-p-k}^\nu\,\widetilde{\bm{\alpha}}_{q-p-k}\cdot\widetilde{\bm{\alpha}}_k+(k+p)\,E_{-k-p}^\nu\,\widetilde{E}_{k+p}^\nu\,E_{k}^\nu\,E_{p-q-k}^\nu\,\widetilde{\bm{\alpha}} _{k}\cdot\bm{\alpha}_{p-q-k}\right\rbrace\,;}&\nonumber
\ee
\be
&\displaystyle{\sum_{n\in\mathbb{Z}}E_{n}^\nu\,\widetilde{E}_{n+q}^\nu\,\left\lbrace L_{p,\nu}(z),\widetilde{\bm{\alpha}}_{n+q}\cdot\bm{\alpha}_{n}\right\rbrace_{P.B.}=-i\sum_{n\in\mathbb{Z}}E_{n}^\nu\,\widetilde{E}_{n+q}^\nu\,\left\lbrace(n+q)\Big[ \underbrace{ E_{n-p+q}^\nu\,E_{-q-n}^\nu\,\widetilde{\bm{\alpha}}_{n-p+q}\cdot\bm{\alpha} _{n}}_{n=k}-\right.}&\nonumber\\
&\displaystyle{\left.- \underbrace{E_{-q-n}^\nu\,\widetilde{E}_{p-q-n}^\nu\,\bm{\alpha} _{p-q-n}\cdot\bm{\alpha} _{n}}_{n=k}\Big]+n\Big[\underbrace{ \widetilde{E}_{n+p}^\nu\,\widetilde{E}_{-n}^\nu\,\bm{\alpha}_{n+p}\cdot\widetilde{\bm{\alpha}}_{n+q}}_{k=n+p}-\underbrace{\widetilde{E}_{-n}^\nu\,E_{-p-n}^\nu\,\widetilde{\bm{\alpha}} _{-p-n}\cdot\widetilde{\bm{\alpha}}_{n+q}}_{-k=p+n}\Big]\right\rbrace\,;}&\nonumber\\
&\displaystyle{=\sum_{k\in\mathbb{Z}}\left\lbrace(k+p)\,E_{-k-p}^\nu\,\widetilde{E}_{p+k}^\nu\,\widetilde{E}_{q-p-k}^\nu\,E_k^\nu\,\widetilde{\bm{\alpha}}_{q-p-k}\cdot\widetilde{\bm{\alpha}}_k+(k+q)\,\widetilde{E}_{k+q}^\nu\,E_{-k-q}^\nu\,E_{k+q-p}^\nu\,E_k^\nu\,\widetilde{\bm{\alpha}}_{k+q-p}\cdot\bm{\alpha}_k-\right.}&\nonumber\\
&\displaystyle{\left.-(k+q)\,\widetilde{E}_{k+q}^\nu\,E_{-k-q}^\nu\,E_k^\nu\, \widetilde{E}_{p-q-k}^\nu\,\bm{\alpha}_{p-q-k}\cdot\bm{\alpha}_k+(k-p)\,E_{k-p}^\nu\,\widetilde{E}_{p-k}^\nu\,\widetilde{E}_{k}^\nu\,\widetilde{E}_{k-p+q}^\nu\,\widetilde{\bm{\alpha}} _{k-p+q}\cdot\bm{\alpha}_{k}\right\rbrace\,.}&\nonumber
\ee
Lastly the final contribution is 
\be
&\displaystyle{\sum_{n\in\mathbb{Z}}\widetilde{E}_{n}^\nu\,\widetilde{E}_{q-n}^\nu\,\left\lbrace L_{p,\nu}(z),\widetilde{\bm{\alpha}}_{q-n}\cdot\widetilde{\bm{\alpha}}_n\right\rbrace_{P.B.}=-i\sum_{n\in\mathbb{Z}}\widetilde{E}_{n}^\nu\,\widetilde{E}_{q-n}^\nu\,\left\lbrace(q-n)\Big[ \underbrace{ E_{q-n-p}^\nu\,E_{n-q}^\nu\,\widetilde{\bm{\alpha}}_{q-p-n}\cdot\widetilde{\bm{\alpha}} _{n}}_{n=k}-\right.
}&\nonumber\\
&\displaystyle{\left.- \underbrace{E_{n-q}^\nu\,\widetilde{E}_{p-q+n}^\nu\,\bm{\alpha} _{p-q+n}\cdot\widetilde{\bm{\alpha}} _{n}}_{n=k}\Big]+n\Big[\underbrace{ E_{n-p}^\nu\,E_{-n}^\nu\,\widetilde{\bm{\alpha}}_{n-p}\cdot\widetilde{\bm{\alpha}}_{-n-q}}_{k=n-p}- \underbrace{\widetilde{E}_{p-n}^\nu\,E_{-n}^\nu\,\bm{\alpha} _{p-n}\cdot\widetilde{\bm{\alpha}}_{q-n}}_{k=p-n}\Big]\right\rbrace\,;}&\nonumber\\
&\displaystyle{=\sum_{k\in\mathbb{Z}}\left\lbrace\Big[(k+p)\,E_{-k-p}^\nu\,\widetilde{E}_{p+k}^\nu\,\widetilde{E}_{q-p-k}^\nu\,E_k^\nu\,-(k-q)\,E_{k-q}^\nu\,\widetilde{E}_{q-k}^\nu\,E_{q-k-p}^\nu\,\widetilde{E}_k^\nu\,\Big]\widetilde{\bm{\alpha}}_{q-k-p}\cdot\widetilde{\bm{\alpha}}_k+\right.
}&\nonumber\\
&\displaystyle{\left.+(k-q)\,E_{k-q}^\nu\,\widetilde{E}_{q-k}^\nu\,\widetilde{E}_k^\nu\,\widetilde{E}_{p-q+k}^\nu\,\bm{\alpha}_{p-q+k}\cdot\widetilde{\bm{\alpha}}_k+\,(k-p)\,E_{k-p}^\nu\,\widetilde{E}_{p-k}^\nu\,\widetilde{E}_{k}^\nu\,\widetilde{E}_{k-p+q}^\nu\,\widetilde{\bm{\alpha}} _{k-p+q}\cdot\bm{\alpha}_{k}\right\rbrace\,.}&\nonumber
\ee
If we add all these contribution we find
\be
\left\lbrace L_{p,\nu}(z),\widetilde{L}_{q,\nu}(z)  \right\rbrace_{P.B.}&=&0\,.
\ee

\section{Asymptotic approximation of the Fractional Virasoro algebra} \label{Appen:AsympViraSums}
In this Appendix we present the asymptotic form of the Fractional Virasoro operators and we explicit computed its algebra.

In the asymptotic limit $z\gg |\nu^2-\frac{1}{4}|$,  the fractional Virasoro Operators take the form
\be
\widetilde{L}_{p,\nu}(z)&=&\frac{1}{2\,\Gamma(2-2\nu)}\sum_{n\in\mathbb{Z}}\ell_{p-n}^\nu(z)\,\ell_n^\nu(z)\,\widetilde{\bm{\alpha}}_{p-n}\cdot\widetilde{\bm{\alpha}}_n\,e^{-ipz}\,;\\
L_{p,\nu}(z)&=&\frac{1}{2\,\Gamma(2-2\nu)}\sum_{n\in\mathbb{Z}}\ell_{p-n}^\nu(z)\,\ell_n^\nu(z)\,\bm{\alpha}_{p-n}\cdot\bm{\alpha}_n\,e^{-ipz}\,,
\ee
where $\ell_{m}^\nu(z)=\exp\Big[-\frac{i\pi}{2}\left(\nu-\frac{1}{2}\right)\left(\texttt{Sgn}(m)+\frac{2i}{\pi}\,\delta_{0,m}\ln(z)\right)\Big]$. Let us compute the Poisson brackets between $L_p$ and the $\alpha's$. All commutators are zero expect for:
\be
\left\lbrace L_{p,\nu}(z),\alpha^\mu _{m}   \right\rbrace_{P.B.}&=& \frac{1}{2\,\Gamma(2-2\nu)}\sum_{n\in\mathbb{Z}}\ell_{p-n}^\nu\,\ell_n^\nu\Big[
\left\lbrace \alpha^\rho _{p-n},  \alpha^\mu _{m} \right\rbrace_{P.B.}\alpha^\rho _{n}+
\alpha^\rho _{p-n}\,\left\lbrace \alpha^\rho _{n},  \alpha^\mu _{m} \right\rbrace_{P.B.}
\Big]\,e^{-ipz}\nonumber\\
&=&-im\,\ell_{-m}^\nu\,\ell_{p+m}^{\nu}\,\alpha_{m+p}^\mu\,e^{-ipz}\,;\\
\left\lbrace \widetilde{L}_{p,\nu}(z),\widetilde{\alpha}^\mu _{m}   \right\rbrace_{P.B.}&=& \frac{1}{2\,\Gamma(2-2\nu)}\sum_{n\in\mathbb{Z}}\ell_{p-n}^\nu\,\ell_n^\nu\Big[
\left\lbrace \widetilde{\alpha}^\rho _{p-n}, \widetilde{\alpha}^\mu _{m} \right\rbrace_{P.B.}\widetilde{\alpha}^\rho _{n}+
\widetilde{\alpha}^\rho _{p-n}\,\left\lbrace \widetilde{\alpha}^\rho _{n},  \widetilde{\alpha}^\mu _{m} \right\rbrace_{P.B.}
\Big]\,e^{-ipz}\nonumber\\
&=&-im\,\ell_{-m}^\nu\,\ell_{p+m}^{\nu}\,\widetilde{\alpha}_{m+p}^\mu\,e^{-ipz}\,.
\ee
Therefore, using these commutators we can compute
\be
\left\lbrace L_{p,\nu}(z),L_{q,\nu}(z)   \right\rbrace_{P.B.}&=&\frac{e^{-iqz}}{2\,\Gamma(2-2\nu)}\sum_{n\in\mathbb{Z}}\ell_{q-n}^\nu\,\ell_n^\nu\,\Big[\left\lbrace L_{p,\nu}(z),\alpha^\rho _{q-n},  \right\rbrace_{P.B.}\alpha^\rho _{n}+\alpha^\rho _{q-n}\left\lbrace L_{p,\nu}(z),\alpha^\rho _{n},  \right\rbrace_{P.B.}\Big]\nonumber\\
&=&-\frac{i\,e^{-i(p+q)z}}{2\,\Gamma(2-2\nu)}\sum_{n\in\mathbb{Z}}\ell_{q-n}^\nu\,\ell_n^\nu\,\Big[(q-n) \,\ell_{n-q}^\nu\,\ell_{p+q-n}^{\nu}\, \bm{\alpha}_{p+q-n}\cdot\bm{\alpha} _{n}+n\,\ell_{-n}^\nu\,\ell_{p+n}^{\nu}\,\bm{\alpha} _{q-n}\cdot \bm{\alpha}_{n+p}\Big]\nonumber
\ee
Setting in the first sum $n=k$ and in the second sum $p+n=k$, we can rewrite the sums as
\be
\left\lbrace L_{p,\nu}(z),L_{q,\nu}(z)   \right\rbrace_{P.B.}&=&-\frac{i\,e^{-i(p+q)z}}{2\,\Gamma(2-2\nu)}\sum_{k\in\mathbb{Z}}\Big[(q-k) \ell_{q-k}^\nu\,\ell_{k-q}^\nu +(k-p)\,\ell_{p-k}^\nu\,\ell_{k-p}^{\nu}\Big]\ell_{p+q-k}^{\nu}\,\ell_k^\nu\,\bm{\alpha}_{p+q-k}\cdot\bm{\alpha} _{k}\nonumber\,.
\ee
We can write the products of $\ell$'s as
\be
\ell_{q-k}^\nu(z)\,\ell_{k-q}^{\nu}(z)&=&e^{2(\nu-\frac{1}{2})\,\ln(z)\,\delta_{0,k-q}}\nonumber\\
&=&1+2\,\sinh(x_{k-q})\,e^{x_{k-q}}\nonumber
\ee
with $x_{k-q}=(\nu-\frac{1}{2})\,\ln(z)\,\delta_{0,k-q}$. Therefore, the commutators of the $L_p(z)$ can be re-written as
\be
\left\lbrace L_{p,\nu}(z),L_{q,\nu}(z)   \right\rbrace_{P.B.}&=&i(p-q)\,L_{p+q,\nu}(z)-i\frac{\left(\nu-\frac{1}{2}\right)}{\Gamma(2-2\nu)}\,\ln(z) \sum_{k\in\mathbb{Z}}\mathcal{O}_k(p,q;z)\,\bm{\alpha}_{p+q-k}\cdot\bm{\alpha} _{k}\nonumber\,,
\ee
where
\be
\mathcal{O}_k(p,q;z)&=&\frac{e^{-i(p+q)z}}{x_0}\,\left[(k-p)\,\sinh(x_{k-p})\,e^{x_{k-p}}-(k-q)\,\sinh(x_{k-q})\,e^{x_{k-q}}\right]\,.
\ee
The study of this function is as follows:
\begin{itemize}
\item \textbf{$\mathcal{O}_k(p,q;z)$ function}.
Let us analyse the function 
$$\mathcal{O}_k(p,q;z)=\frac{e^{-i(p+q)z}}{x_0}\,\left[(k-p)\,\sinh(x_{k-p})\,e^{x_{k-p}}-(k-q)\,\sinh(x_{k-q})\,e^{x_{k-q}}\right]\,.$$ 
It is clear that $\mathcal{O}_k(p,p;z)=\mathcal{O}_k(q,q;z)=0$. Special cases of this function are, for $p\neq q$, when $k=p$ and $k\neq q$, $k=q$ and $k\neq p$, and finally $k\neq p,q$.
\begin{itemize}
\item
\textbf{Case $k=p$ and $k\neq q$}.
\be
\mathcal{O}_p(p,q;z)&=&\frac{e^{-i(p+q)z}}{x_0}\,\left[-p\,\sinh(x_{0})\,e^{x_{0}}+p\left(\sinh(x_{0})\,e^{x_{0}}\right)\right]=0\,.\nonumber
\ee
\item
\textbf{Case $k=q$ and $k\neq p$}.
\be
\mathcal{O}_q(p,q;z)&=&\frac{e^{-i(p+q)z}}{x_0}\,\left[q\,\sinh(x_{0})\,e^{x_{0}}-q\left(\sinh(x_{0})\,e^{x_{0}}\right)\right]=0\,.\nonumber
\ee

\item
\textbf{Case $k\neq q$ and $k\neq p$}.
\be
\mathcal{O}_k(p,q;z)&=&\frac{e^{-i(p+q)z}}{x_0}\,\left[(k-p)\,\sinh(x_{k-p})\,e^{x_{k-p}}-(k-q)\,\sinh(x_{k-q})\,e^{x_{k-q}}\right]=0\,,\nonumber
\ee
because $x_{k-p}=x_{k-q}=0$.
\end{itemize}
Same procedure can be repeated for $\mathcal{O}_k(p,-p;z)$. Therefore, we can conclude that $\mathcal{O}_k(p,q;z)=0$ for all $p$, $q$, $k$ and $z$.
\end{itemize}

Finally, we can write
\be
\left\lbrace L_{p,\nu}(z),L_{q,\nu}(z)   \right\rbrace_{P.B.}&=&i(p-q)\,L_{p+q,\nu}(z)\nonumber\,.
\ee
For the $\widetilde{L}$ the computation is completely analogue.

\end{document}